\documentclass[preprint,12pt]{elsarticle_mod}

\newcommand{\rr}[1]{{\normalfont\textrm{#1}}}

\newlength{\pecettawidth}
\usepackage{amssymb}

\begin{document}

\begin{frontmatter}



\title{Dynamics of pedestrians in regions with no visibility -- a lattice model without exclusion}


\author{Emilio N.\ M.\ Cirillo}
\ead{emilio.cirillo@uniroma1.it}
\author{Adrian Muntean}
\address{Dipartimento di Scienze di Base e Applicate per 
             l'Ingegneria, Sapienza Universit\`a di Roma, 
             via A.\ Scarpa 16, I--00161}
\address{Department of Mathematics and Computer Science, (CASA) Centre for Analysis, Scientific computing and Applications, Institute for Complex Molecular Systems (ICMS)
             Eindhoven University of Technology,
             P.O.\ Box 513, 5600 MB Eindhoven, The Netherlands}
\ead{a.muntean@tue.nl}

\begin{abstract}
We investigate the motion of pedestrians through obscure corridors where the lack of visibility (due to smoke, fog, darkness, etc.) hides the precise position of the exits. 
We focus our attention on a set of basic mechanisms, which we assume to be governing the dynamics at the individual level. Using a lattice model, we explore the effects of 
non--exclusion  on the overall exit flux (evacuation rate). More precisely, we study the effect of the buddying threshold 
(of no--exclusion per site) on the dynamics of the crowd and investigate to which extent  our model confirms the following pattern  revealed by investigations on real emergencies: If the evacuees tend to cooperate and act altruistically, then their collective action tends to  favor the occurrence of disasters. 
\end{abstract}


\end{frontmatter}


\section{Introduction}\label{s:int}
\par\noindent

Efficient evacuation of humans from high--risk zones is a very important societal issue \cite{Armin}. The topic is very well studied by large communities of scientists ranging from logistics and transportation, civil and fire engineering, to theoretical physics or applied mathematics.  Many modeling approaches  (deterministic or stochastic) succeed to capture qualitatively basic behaviors of humans (here referred to as pedestrians ) walking within a given geometry towards  {\em a priori} prescribed exits. For instance,  social force/social velocity crowd dynamics models (cf. e.g.  \cite{HelbingMolnar}, \cite{PiccoliTosinMeasTh}, \cite{EversMuntean}), simple asymmetric exclusion models (see chapters 3 and 4 from \cite{Schadschneider2011} as well as references cited therein), cellular automaton-type models \cite{Kirchner,Guo}, etc.\footnote{For a detailed classification of pedestrian models, see \cite{Schadschneider2011}.} are quite efficient to evacuate people when exits are visible.

The driving question of our research is: {\em What to do when exits are not visible?}  A search through the existing literature shows that there are studies done [especially for fire evacuation scenarios] on how information and way finding systems are perceived by individuals.  One of the main questions in fire safety research is whether green flashing lights can influence the evacuation (particularly, the exit choice); see e.g. \cite{Nilsson,Shields,Bauke} (and the fire engineering references cited therein) and \cite{Yuan} (partial visibility due to a 
non--uniform smoke concentration) \cite{Jun} (partial visibility as a function of smoke's temperature), \cite{Zheng} (flow heterogeneity due to fire spreading). On the other hand, as far as we are aware of, there is no literature on dynamics of pedestrians (or groups of pedestrians) in regions with no visibility. 

 In this paper, we aim at investigating the motion of pedestrians through obscure corridors where the lack of visibility (due to smoke, fog, darkness, etc.) hides the precise position of the exits. We focus our attention on a set of basic mechanisms, which we assume to be governing the dynamics at individual level. Using a lattice model, we explore the effects of 
non--exclusion  on the overall exit flux (evacuation rate). More precisely, we study the effect of the buddying threshold 
(of no--exclusion per site) on the dynamics of the crowd and wish to investigate to which extent  our model confirms the following pattern  revealed by investigations on real emergencies: If the evacuees tend to cooperate and act altruistically, then their collective action tends to  favor the occurrence of disasters.\footnote{Note that due to the lack of visibility anticipation effects (see \cite{Suma}) and drifts (see \cite{Guo}) are expected to play no role in evacuation.}

By means of a minimal model, we wish to describe how a bunch of people located inside a dark (smoky, foggy, etc.) corridor  exits through 
an invisible door open in one of the four walls. A few central driving questions need to be mentioned at this point: 
\begin{center} 
\begin{itemize}
\item[(Q1)] {\em How do pedestrians choose their path and speed when they are about  to move through regions with no visibility? }
\item[(Q2)] {\em Is group formation (e.g. buddying) the right strategy to move through such uncomfortable zones  able to ensure exiting  
within a reasonable time?}
\end{itemize}
\end{center}
It is worth noting that this is one of the situations where group psychology (compare e.g.  \cite{LeBon,Curseu} and \cite{DyerJohanssonHelbing}) is not really helping, especially due to the lack of experimental observations. On the other hand, the concept of group we look at here is not similar to the swarming patterns typically emphasized in nature by fish and or birds communities 
(see e.g. the 4--groups taxonomy in  \cite{Topaz}, namely swarm, torus, dynamic parallel groups, and highly parallel groups). On top of this, note that concepts like leadership (see e.g. \cite{CouzinNature}) or motion fluidization by favorizing lanes formation (see e.g. \cite{HelbingVicsek})  are simply not applicable in this context.

The reminder of the paper has the following structure: In section \ref{s:modello} we introduce our model. Section \ref{s:risultati} presents our simulation results, while we close with comments, discussions and suggestions for further investigations in Section \ref{s:discussion}.

\section{A lattice model for  the reverse {\em mosca cieca} game}
\label{s:modello}
We make here first attempts in answering the questions (Q1) and (Q2). 
To do this end we employ a minimal lattice model, named the 
{\em reverse mosca cieca game}\footnote{{\em Mosca cieca} 
  means in Italian {\em blind fly}.  It  
 is the Italian name of 
 a traditional children's game also known as 
 {\em blind man's buff} or {\em blind man's bluff}.
 The game is played in a spacious free of dangers area, such as a large room, 
 in which one player, the ``mosca", is blindfolded and moves around 
 attempting to catch the other players without being able to see them, 
 while the other players try to avoid him, hiding in plain sight and 
 making fun of the ``mosca" to make him/her change direction.
 When one of the player is finally cautgh,  the ``mosca" has to identify 
 him by touching is face 
 and if the person is correctly identified he becomes the ``mosca".
 Our model is a sort of reverse mosca cieca game since all the 
 players (the pedestrians) group around, as if they were
 blindfolded, trying to catch the (invisible) exit.
},
where we incorporate a few basic rules for the pedestrians motion.

\par\noindent
\subsection{Basic assumptions on the pedestrians motion}
\par\noindent
We assume that the motion of pedestrians is 
governed by the following four mechanisms:
\begin{itemize}
\item[(A1)]
In the core of the corridor, 
people move freely without constraints;
\item[(A2)]
The boundary is reflecting, possibly attracting;
\item[(A3)]
People are attracted by bunches of other people up 
to a threshold;
\item[(A4)] 
People are blind in the sense that there is no drift (desired velocity) leading them 
towards the exit.
\end{itemize}

(A1)--(A4) intend to describe the following situation: 

Since, in this framework, neighbors (both individuals or groups) can not be visually identified by the individuals in motion, basic mechanisms like attraction to a group, tendency to align, or social repulsion are negligible and individuals have to live with ``preferences". 
Essentially, they move freely inside the corridor but they like to buddy to people they accidentally meet at a certain point (site). The more people are localized at a certain site, the stronger the preference to attach to it.  However if the number of people at a site reaches a threshold, then such site becomes 
not attracting 
for eventually new incomers. We refer to (A3) as the {\em buddying mechanism}. 

Once an individual touches a wall, he/she simply felts the need to stick to it at least for a while, i.e. until he/she  can attach to an interesting site (having conveniently many hosts) or to a group of unevenly occupied sites or the exact location of the door is detected. 

Since people have no desired velocity, their diffusion (random walk) together with the buddying are the only transport mechanisms. Can these  eventually lead to evacuation?

In the following, we study the effect of the threshold 
(of no--exclusion per site) 
on the overall dynamics of the crowd. We will describe our results in terms of macroscopic quantities like averaged outgoing flux, stationary occupation numbers,  and  stationary correlations.

\subsection{The lattice model}
\par\noindent
To describe our model, we start off with the construction of the lattice.  
It is worth mentioning that the 
approach we take here is very much influenced by a basic scenario described in \cite{Emilio} for randomly moving sodium ions willing to pass through a switching 
on--off  membrane gate. The major difference here is twofold: the gate is permanently open and the buddying principle is activated.

Let $e_1:=(1,0)$ and $e_2=(0,1)$ denote the coordinate vectors in $\mathbb{R}^2$.
 Let $\Lambda\subset\mathbb{Z}^2$ be a finite square with 
odd side length $L$. We refer to this as  the {\em corridor}. 
Each element $x$ of $\Lambda$ will be called a {\em cell} or {\em site}.
The external boundary of the corridor is made of four segments 
made of $L$ cells each; 
the point at the center of one of these four sides is called {\em exit}.

Let $N$ be  positive integer denoting the (total) {\em number of individuals} inside 
the corridor $\Lambda$. 
We consider the state space $X:=\{0,\dots,N\}^\Lambda$.  
 For any state $n\in X$, we let 
$n(x)$ be the {\em number of 
individuals} at cell $x$\footnote{Note that, at each cell $x$, we don't control how many individuals are located. We only know that $n(x)\in [0,N]$ for all $x\in \Lambda$.}.

We define a Markov chain $n_t$ on the finite state space 
$X$ with discrete time $t=0,1,\dots$.
The parameters of the process will be the integers (possibly equal 
to zero) $Q,T,W\ge0$ called respectively 
{\em minimal quantum}, {\em threshold}, and {\em wall interaction},
and the real number $R\in[0,1]$ called the {\em rest parameter}.
We finally define the function $S:\mathbb{N}\to\mathbb{N}$ 
such that 
\begin{displaymath}
S(k):=
\left\{
\begin{array}{ll}
Q & \textrm{ if } k>T\\
k+Q & \textrm{ if } k\le T\\
\end{array}
\right.
\end{displaymath}
for any $k\in\mathbb{N}$. Note that for $k=0$ we have $S(0)=Q$.

At each time $t$, the $N$ individuals move simultaneously within the corridor 
according to the following rule:

For any cell $x$ situated in the interior of the corridor $\Lambda$, and all 
$y$ nearest neighbor of $x$,
with  $n\in X$, we define the weights
\begin{displaymath}
w(x,x):=RS(n(x))
\;\;\textrm{ and }\;\;
w(x,y):=S(n(y)).
\end{displaymath}
Also, we obtain the associated probabilities 
$$p(x,x) \mbox{ and } p(x,y)$$ 
by dividing the weight by the normalization 
\begin{displaymath}
w(x,x)+
\sum_{i=1}^2 w(x,x+e_i)+
\sum_{i=1}^2 w(x,x-e_i).
\end{displaymath}
Let now $x$ be in one of the four corners of the corridor $\Lambda$,  and take
$y$ as one of the two nearest neighbors of $x$ inside $\Lambda$. For $n\in X$, we define the weights
\begin{displaymath}
w(x,x):=R(S(n(x))+2W)
\textrm{ and }
w(x,y):=S(n(y))+W
\end{displaymath}
and the associated probabilities $p(x,x)$ and $p(x,y)$ obtained 
by dividing the weight by the suitable\footnote{To get a probability, we divide the weights
by  the sum of the involved weights: It is about 3 terms if the 
site lies in a corner of the corridor, 4 terms if it is close to a border, and then 
5 terms for a pedestrian located in the core. } normalization. 

It is worth stressing here that $T$ is not a threshold in $n(x)$ -- the number of individuals per cell. It is a threshold in the probability that such a cell is likely to be occupied or not.  
 
For $x\in \Lambda$ neighboring the boundary (but neither in the corners, nor  
neighboring the exit), 
$y$ one of the two nearest neighbor of $x$ inside $\Lambda$ and 
neighboring the boundary,
$z$ the nearest neighbor of $x$ in the interior of $\Lambda$,
and $n\in X$, we define the weights
\begin{displaymath}
\begin{array}{l}
w(x,x):=R(S(n(x))+W)\\
w(x,y):=S(n(y))+W\\
w(x,z):=S(n(y)).\\
\end{array}
\end{displaymath}
The associated probabilities $p(x,x)$, $p(x,y)$, and $p(x,z)$ are obtained 
by dividing the weight by the suitable normalization. 

Finally, if $x$ is the cell in $\Lambda$ neighboring the exit and 
$y$ is one of the two nearest neighbor of $x$ inside $\Lambda$ and 
neighboring the boundary,
$z$ being the nearest neighbor of $x$ in the interior of $\Lambda$,
and $n\in X$, we define the weights
\begin{displaymath}
\begin{array}{l}
w(x,x):=RS(n(x))\\
w(x,y):=S(n(y))+W\\
w(x,z):=S(n(y))\\
w(x,\textrm{exit}):=T+Q.\\
\end{array}
\end{displaymath}
The associated probabilities $p(x,x)$, $p(x,y)$, $p(x,z)$,
and $p(x,\textrm{exit})$ are obtained 
by dividing the weight by the suitable normalization. 

The dynamics is then defined as follows:

At each time $t$,  the position of all the individuals 
on each cell is updated according to the probabilities defined 
above. If one of the individuals jumps on the exit 
cell a new individual is put  on a cell of $\Lambda$ 
chosen randomly with the uniform probability $1/L^2$. 

\section{Interpretation of the buddying threshold $T$}
\label{s:commenti}
%
%
%

The possible choices for the parameter $T$ correspond to two different physical 
situations. The first one, for $T=0$,  the function $S(k)$ is equal to $q$ (the minimal 
quantum) whatever the occupation numbers\footnote{The definition of occupation numbers is given in  section~\ref{s:occupo}.} are. This means that
each individual has the same probability to jump to one of its nearest neighbors
or to stay on his site. This is resembling  the independent
symmetric random walk case; the only difference is that with the same probability 
the individuals can decide not to move. We expect that this ``rest probability" just changes a little 
bit the time scale. 

The second physical case is $T>0$. For instance,  $T=1$ means mild buddying, while 
$T=100$ would express an extreme buddying. No simple exclusion is included in this model:
on each site one can cluster as many particles (pedestrians) as one wants. 
The basic role of the threshold is the following: 
The weight associated to the jump towards the site $x$ 
increases from $Q$ to $Q+T$ proportionally to the occupation 
number $n(x)$ until $n(x)=T$, after that level it drops back to $Q$. 
Note that this rule is given on weights and not to probabilities. Therefore,  
if one has $T$ particles at $y$ and $T$ at each of its nearest 
neighbors, then  at the very end one will have that the probability to stay 
or to jump to any of the nearest neighbors is the same. Differences 
in probability are seen only if one of the five (sitting in the core) sites 
involved in the jump (or some of them) has an occupation number 
large (but smaller than the threshold). 
 
\begin{figure}[htbp]
\begin{center}
\begin{picture}(500,200)
\put(-60,160)
{
\resizebox{8cm}{!}{\rotatebox{-90}{\includegraphics{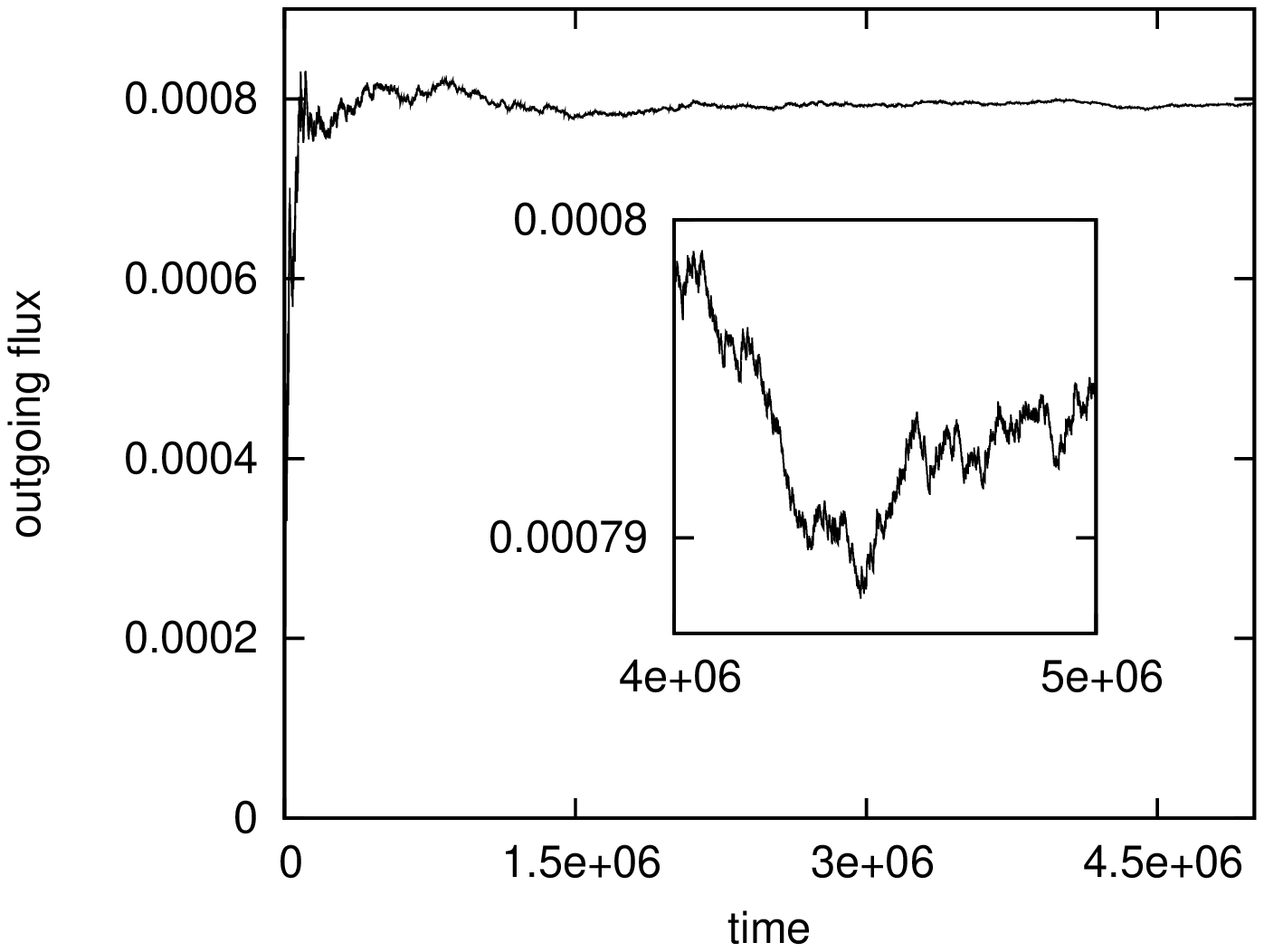}}}
}
\put(180,160)
{
\resizebox{8cm}{!}{\rotatebox{-90}{\includegraphics{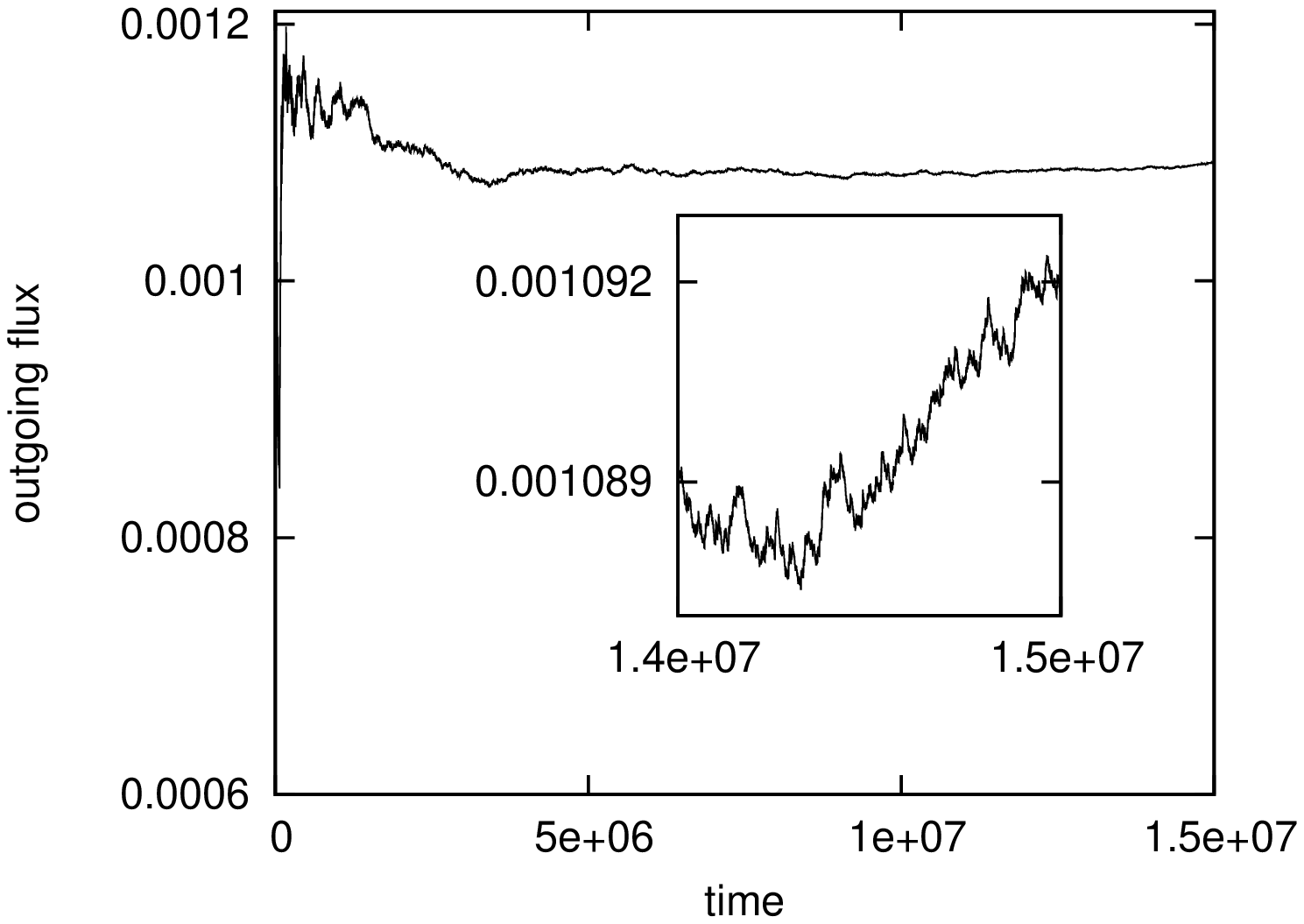}}}
}
\end{picture}
\caption{Averaged outgoing flux vs. time in the case 
$T=0$ and $N=100$ on the left and 
$T=100$ and $N=100$ on the right. 
The inset is a zoom in the time interval 
$[4\times10^6,5\times10^6]$ on the left and 
$[1.4\times10^7,1.5\times10^7]$ on the right.}
\label{f:uno}
\end{center}
\end{figure}

\section{Results}
\label{s:risultati}
\par\noindent
We have studied numerically the model described in section~\ref{s:modello} 
in the case of purely reflecting boundary and eventually resting 
individuals, that is we have considered the case
\begin{displaymath} 
R=1
\;\;\;
\textrm{ and }
\;\;\;
W=0.
\end{displaymath} 
We have also fixed to one the value of the minimal quantum and 
to $101$ the length of the side of the square $\Lambda$, 
that is 
\begin{displaymath} 
Q=1
\;\;\;
\textrm{ and }
\;\;\;
L=101.
\end{displaymath} 
We have studied the behavior of the system by varying the 
treshold parameter, in particular we have considered the cases 
\begin{displaymath}
T=0,1,5,30,100.
\end{displaymath}
It is worth noting  that in the case $T=0$ no buddying effect is introduced 
into the model, that is the individuals behave as standard 
independent random walkers. 
The number of individuals has been also varied; for each value of the 
treshold we have studied the cases
\begin{displaymath}
N=100,600,1000,6000,10000.
\end{displaymath}
For the peculiar values $T=30,100$ more cases have been taken into account.
More precisely,  for the choices $T=30$ and $T=100$ we have also analyzed the supplementary 
cases 
\begin{displaymath}
N=2000,2200,2400,2600,2800,3000,3300
\end{displaymath}
and
\begin{displaymath}
N=1300,1600,2000,3000,
\end{displaymath}
respectively.

\par\noindent
\subsection{Average outgoing flux}
\label{s:flusso}
\par\noindent
The first interesting quantity that one has to compute 
is obviously the {\em average outgoing flux} 
that is to say the ratio between the number of 
individuals which exited the corridor in the time interval $[0, t_\rr{f}]$ 
and $t_\rr{f}$.

This quantity fluctuates in time, 
but for times large enough it approaches a 
constant value. In order to observe relative fluctuations smaller than 
$10^{-2}$ we had to use $t_\rr{f}=5\times10^{6}$. To capture  the extreme
buddying case $T=100$, we used $t_\rr{f}=1.5\times10^{7}$ 
(see figure~\ref{f:uno}).

\begin{figure}[htbp]
\begin{center}
\includegraphics[angle=-90,width=0.75\textwidth]{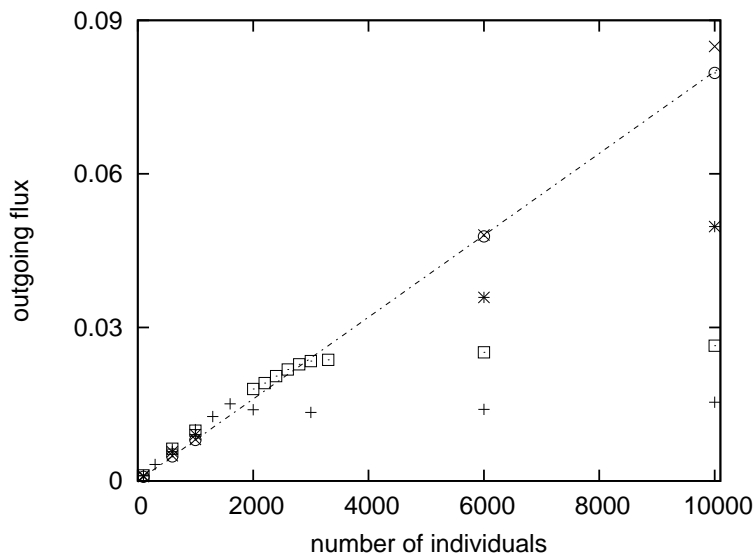}\\
\caption{Averaged outgoing flux vs. number of pedestrians.  
The symbols $\circ$, $\times$, $*$, $\square$, and $+$ 
refer respectively to the cases $T=0,1,5,30,100$.
The straight line has slope $8\times10^{-6}$ and has 
been obtained by fitting the Monte Carlo 
data corresponding to the case $T=0$.}
\label{f:due}
\end{center}
\end{figure}

Figure~\ref{f:due} depicts our results, where the averaged outgoing flux 
is given as a function of the 
number of individuals. 
At $T=0$, that is when no buddying between the individuals is 
put into the model, the outgoing flux results proportional to 
the number of pedestrians in the corridor;
indeed the data represented by the symbol $\circ$ in Figure~\ref{f:due}
have been perfectly fitted by a straight line. 

The appearance of the straight line was expected in the case $T=0$ since 
in this case the dynamics reduces to that of a simple symmetric 
random walk with reflecting boundary conditions. This effect was 
studied rigorously in the one--dimensional case and via Monte Carlo 
simulations in dimension two in \cite{Emilio}. The order of magnitude of the 
slope can be guessed with a simple argument: 
in two dimension the typical time for a walker started to any 
point in $\Lambda$ to reach the boundary is $L^2$ and the number of 
times the walker has to touch the boundary to reach the exit is $L$. 
So that the typical time for a walker to reach the exit is $L^3$. 
This argument yields that the outgoing flux is of order $N/L^3$.
Since in the simulation we used $L=101$, we get an estimate 
of the slope approximatively 
eight time smaller with respect to the measured one. 

When a weak buddying effect is introduced in the model, that is in the 
case $T=1$, we find that if the number of individuals is small enough, 
say 
$N\le6000$, the behavior is similar to the one measured in the absence of 
buddying ($T=0$). At $N=10000$, on the other hand, 
we measure a larger flux; meaning that in the {\em crowded} regime 
small buddying favors the evacuation of the corridor [i.e. it favors the finding of the door]. 

The picture changes completely when buddying is increased. To this end, 
see the cases $T=5,30,100$. 
The outgoing flux is slightly favored when the number of individuals 
is low and strongly depressed when it this becomes high. 
The value of $N$ at which this behavior changes strongly depends 
on the threshold parameter $T$. 

\par\noindent
\subsection{Stationary occupation number}
\label{s:occupo}
\par\noindent
In order to have a deeper insight in the behavior of the system 
we have computed the so called {\em stationary occupation numbers}.
We let $\langle\cdot\rangle$ be the 
stationary average of a random variable associated with the system
and $u(x)$, for any $x\in\Lambda$, 
the random variable giving 
the number of particles $n(x)$ on the site $x$ divided 
times the density $N/L^2$ (with this normalization we expect to get 
typical values close to one). 
We define 
{\em stationary occupation number} at site $x$ as the stationary 
mean $\langle u(x)\rangle$. 

From the computational point of view this task is not an easy one 
for two different reasons:
\begin{itemize}
\item[--]
The presence of the exit in the middle of one of the four 
boundaries makes the system not translationally invariant. Therefore,  
we expect that the occupation number $\langle u(x)\rangle$ does depend on the site $x$;
\item[--]
Computing $\langle u(x)\rangle$ amounts to give a Monte Carlo estimate of the 
stationary measure of the Markov Chain. 
\end{itemize}

In order to estimate the occupation number, we have run jobs 
of length $t_\rr{f}=3\times10^7$ and measured the occupation 
number each $t_\rr{meas}=100$ Monte Carlo steps after having 
waited $t_\rr{term}=100000$ to let the system termalize and 
loose the memory of the starting condition.

The parameter $t_\rr{meas}$ was chosen equal to $100$ after 
analyzing the autocorrelation functions. 
Given a sequence of data $m(j)$ with $j=1,\dots,J$ 
obtained at times $j$, the autocorrelation function \cite{NB} is defined as 
\begin{displaymath}
a(\ell)
:=
\frac{1}{\rr{Var}}
\Big[
\frac{1}{J}
\sum_{j=1}^{J}m(j)m(j+\ell)-
\Big(\frac{1}{J}\sum_{j=1}^{J}m(j)\Big)^2
\Big]
\end{displaymath}
for $\ell=0,\dots,J'$  with $J'\ll J$, where $\rr{Var}$ is the 
variance of the sample defined as 
\begin{displaymath}
\rr{Var}
:=
\frac{1}{J}
\sum_{j=1}^{J}(m(j))^2-
\Big(\frac{1}{J}\sum_{j=1}^{J}m(j)\Big)^2
\end{displaymath}
Autocorrelations are supposed to decay exponentially, so that 
the {\em autocorrelation time} is defined as the smallest time 
such that the autocorrelation is smaller than $1/e$. 
When stationary observables are computed via a Monte Carlo scheme, 
one constructs averages by sampling data separated in time 
by at least one correlation time, preferably two.

\begin{figure}[htbp]
\begin{center}
\begin{picture}(500,400)
\put(0,160)
{
\resizebox{6.5cm}{!}{\rotatebox{-90}{\includegraphics{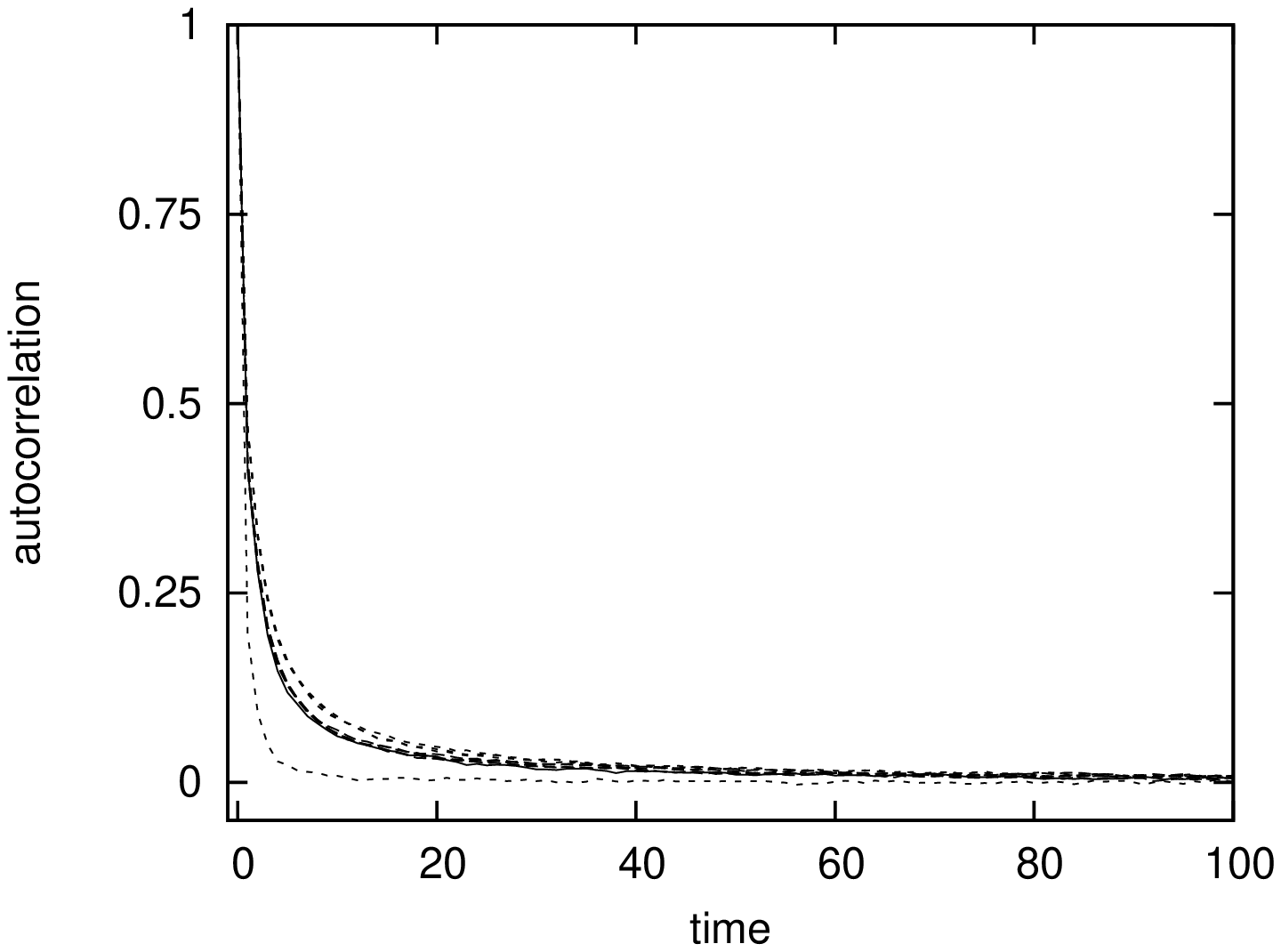}}}
}
\put(180,160)
{
\resizebox{6.5cm}{!}{\rotatebox{-90}{\includegraphics{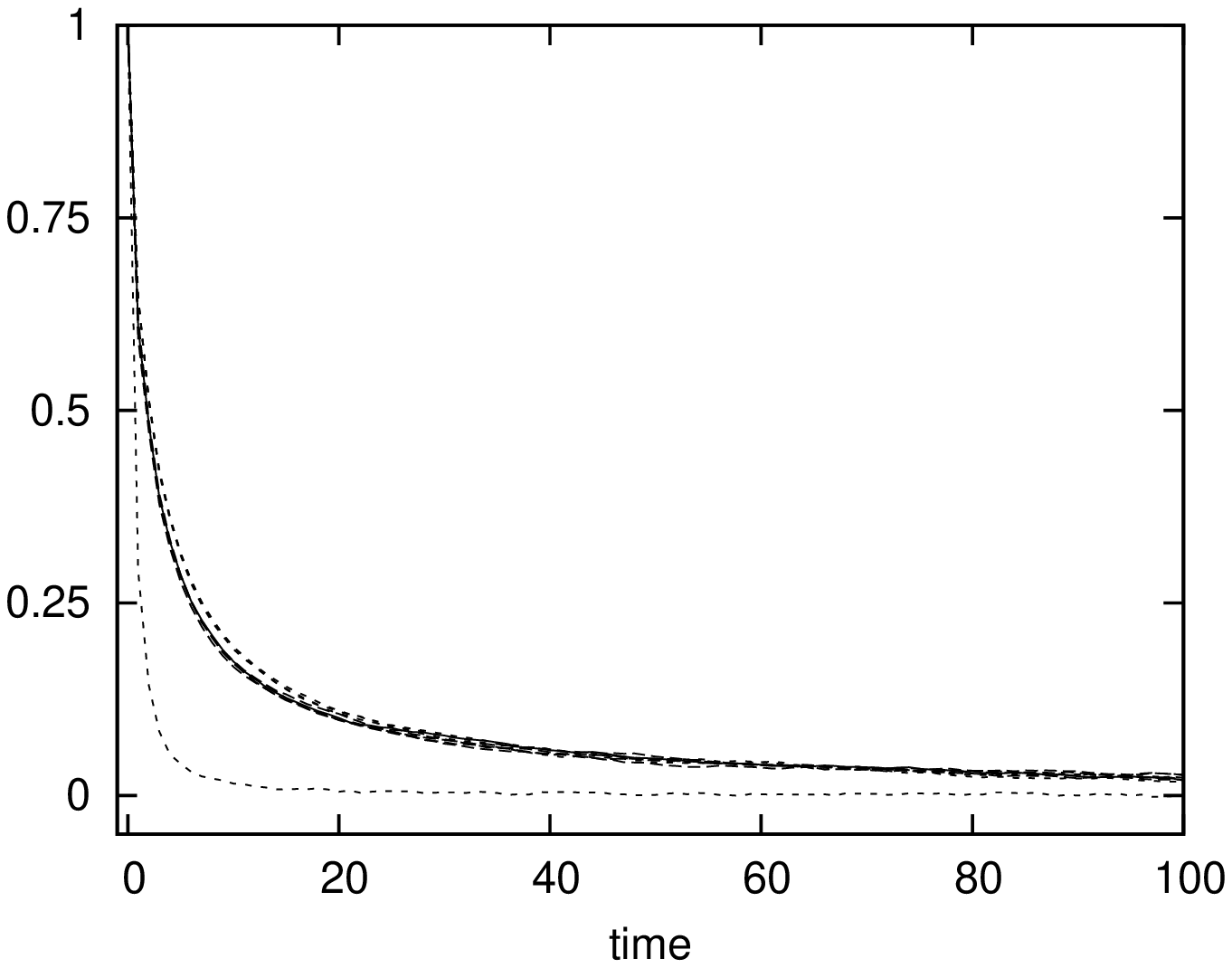}}}
}
\put(0,300)
{
\resizebox{6.5cm}{!}{\rotatebox{-90}{\includegraphics{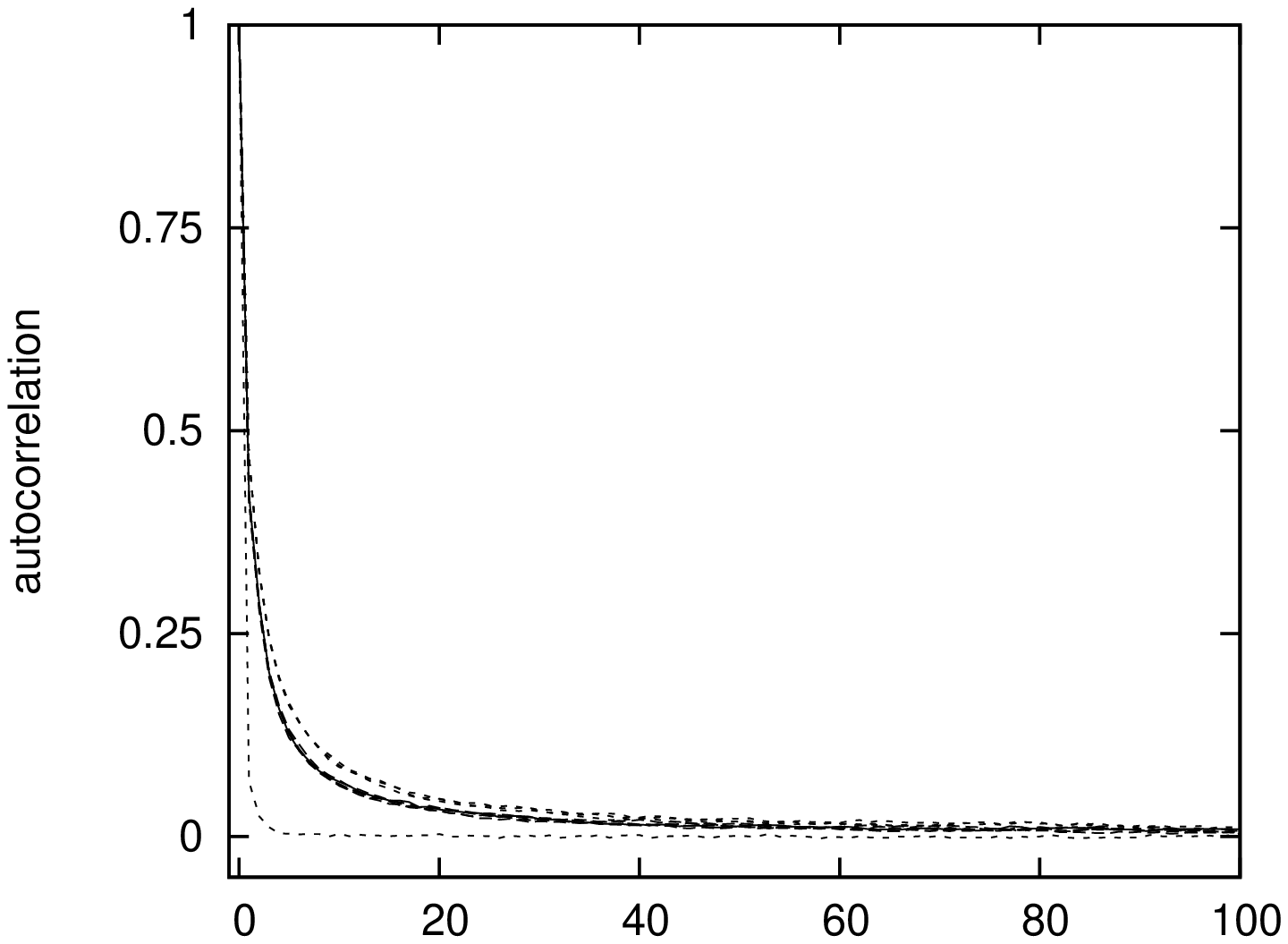}}}
}
\put(180,300)
{
\resizebox{6.5cm}{!}{\rotatebox{-90}{\includegraphics{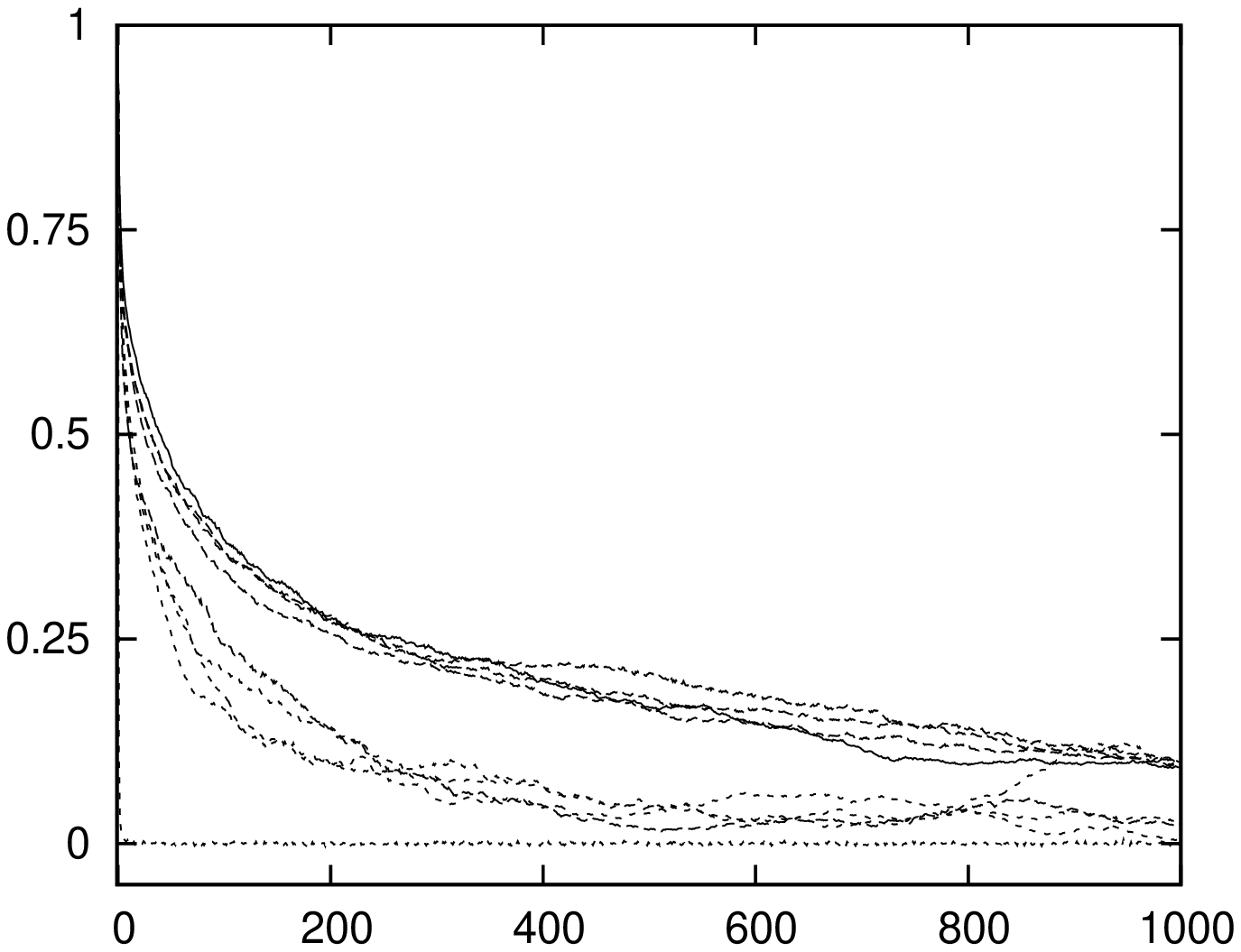}}}
}
\put(0,440)
{
\resizebox{6.5cm}{!}{\rotatebox{-90}{\includegraphics{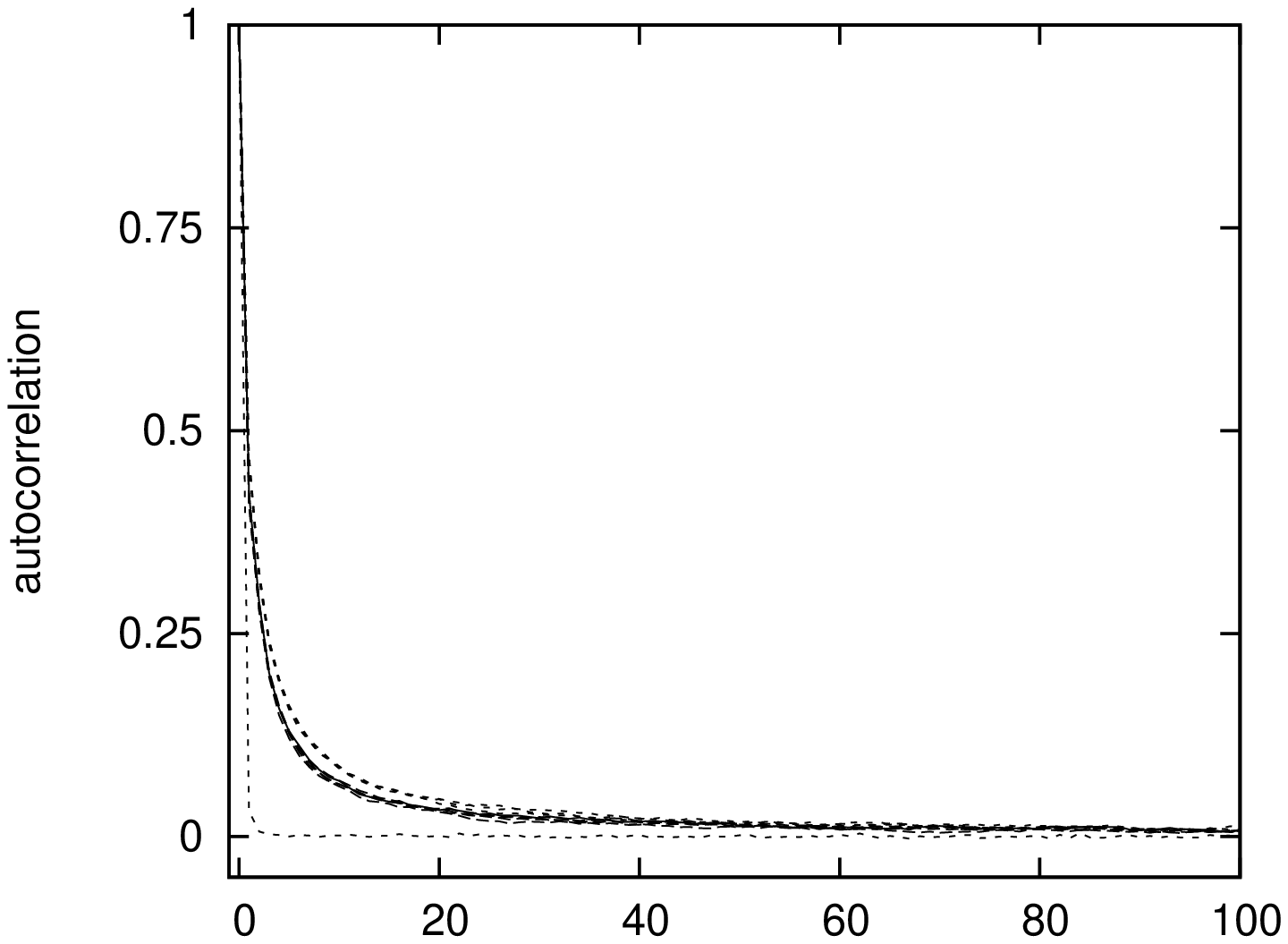}}}
}
\put(180,440)
{
\resizebox{6.5cm}{!}{\rotatebox{-90}{\includegraphics{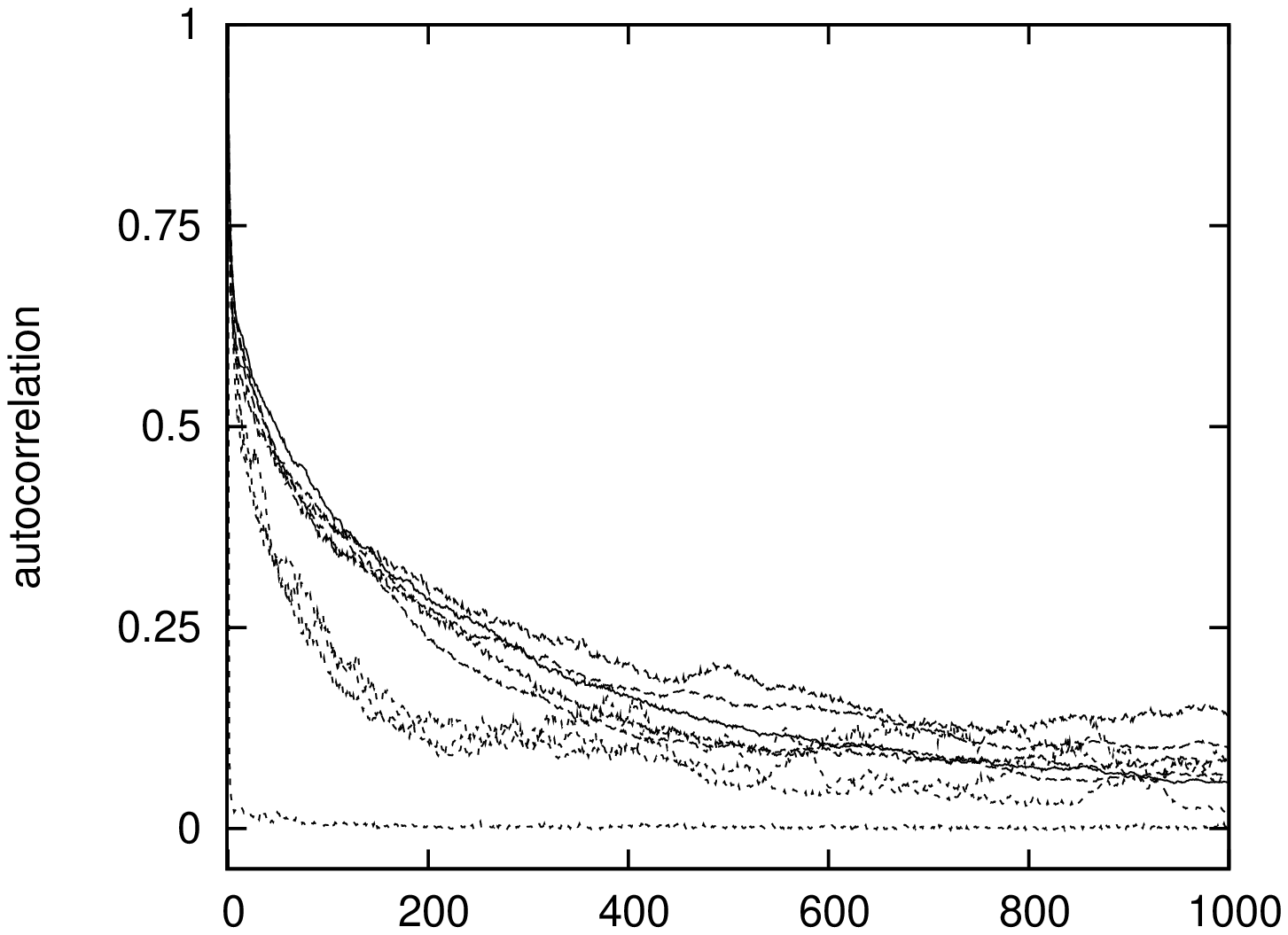}}}
}
\end{picture}
\caption{Autocorrelation vs. time for the 
occupation number at the sites described in the text.
Plots refer to the case $N=1000,10000$ from the left to the 
right and $T=5,30,100$ from the bottom to the top. Solid, dashed, 
and dotted lines refer respectively to the center of the lattice, 
to sites at distance $\lfloor L/4\rfloor$ from the center, and 
to sites at distance $\lfloor L/2\rfloor$ from the center.}
\label{f:tre}
\end{center}
\end{figure}

In figure~\ref{f:tre} we have depicted the autocorrelation 
of the occupation number computed at the center of the lattice,
at the four sites on the axes parallel to the coordinates ones 
and passing through the center of the corridor at distance 
$\lfloor L/4\rfloor$ from the center itself, 
and 
at the four sites on the axes parallel to the coordinates ones 
and passing through the center of the corridor at distance 
$\lfloor L/2\rfloor$ from the center itself. We have 
depicted the data in the cases $T=5,30,100$ and $N=1000,10000$. 
In all the cases the correlation time is smaller than $100$. So that 
chosing $t_\rr{meas}=100$ seems to be a reasonable choice. 

\begin{figure}[htbp]
\begin{center}
\begin{picture}(500,400)
\put(-80,160)
{
\resizebox{6.5cm}{!}{\rotatebox{-90}{\includegraphics{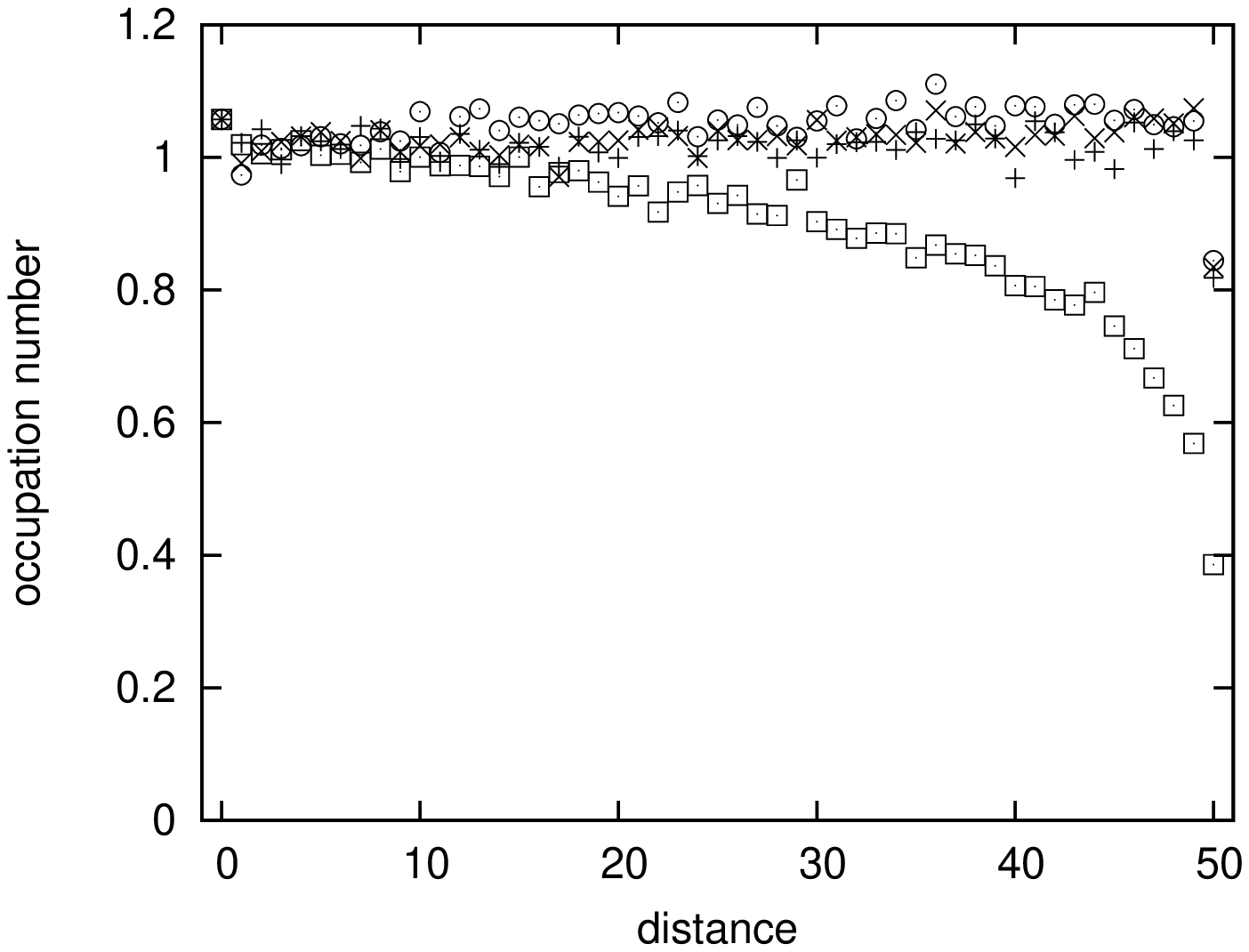}}}
}
\put(90,160)
{
\resizebox{6.5cm}{!}{\rotatebox{-90}{\includegraphics{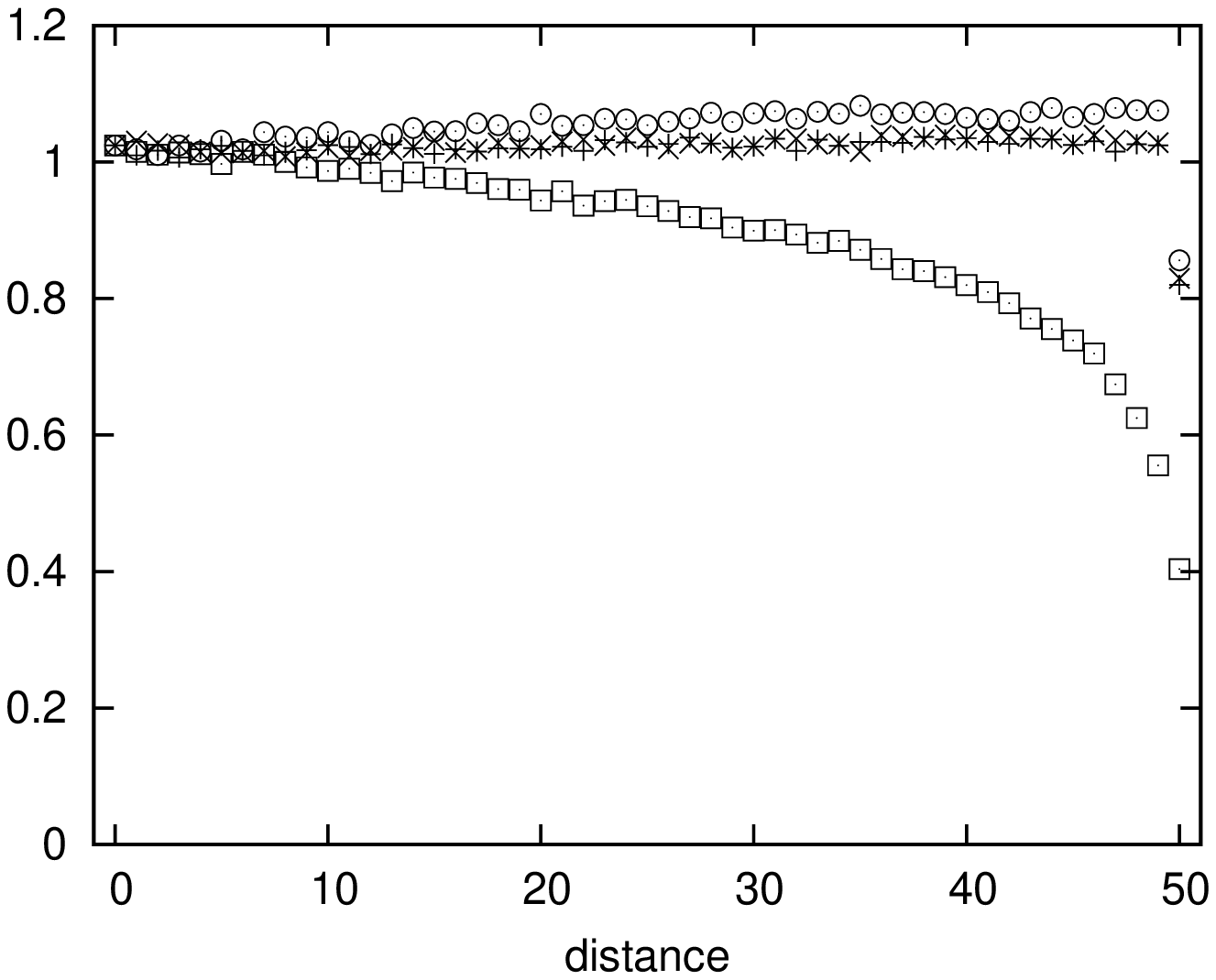}}}
}
\put(260,160)
{
\resizebox{6.5cm}{!}{\rotatebox{-90}{\includegraphics{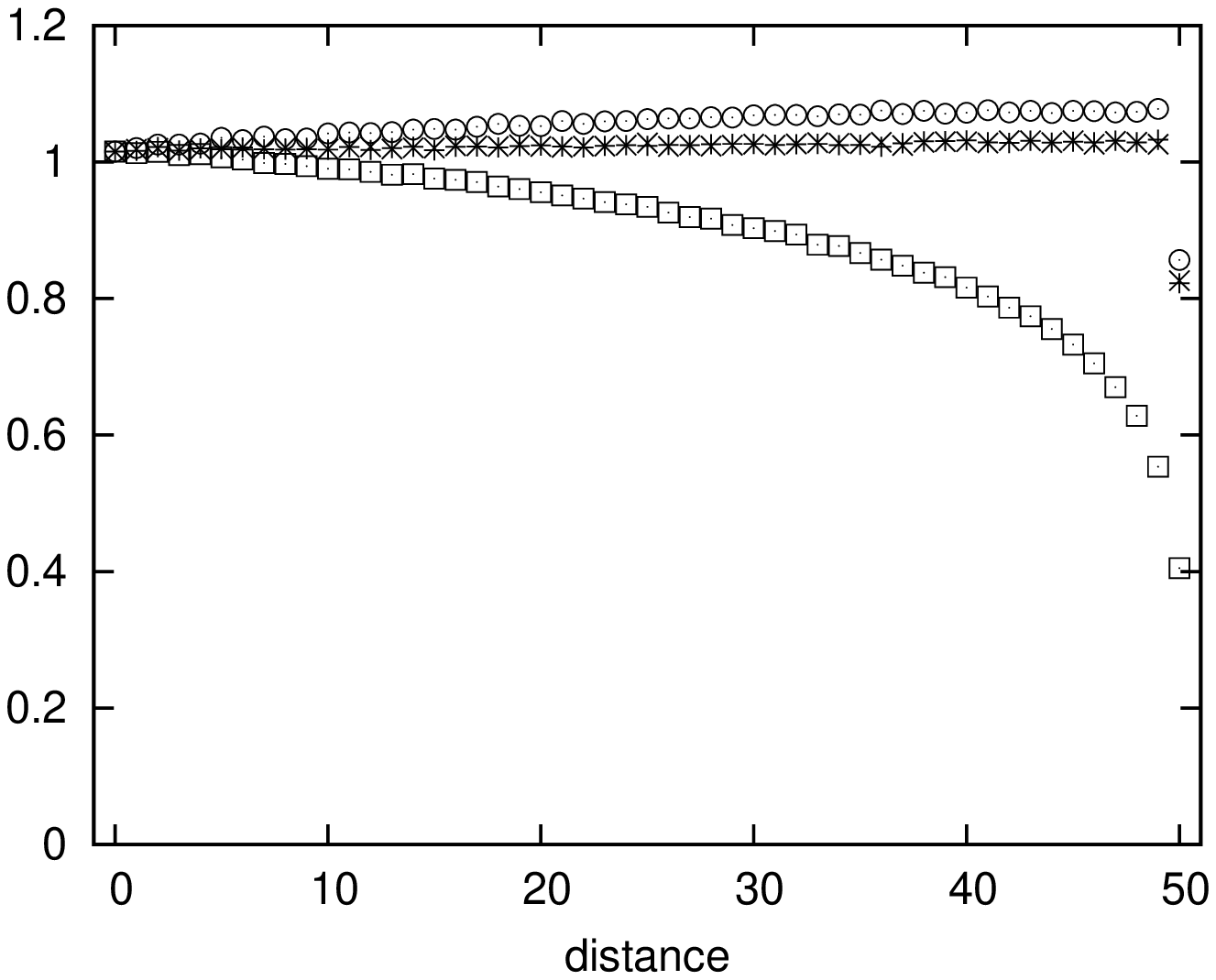}}}
}
\put(-80,300)
{
\resizebox{6.5cm}{!}{\rotatebox{-90}{\includegraphics{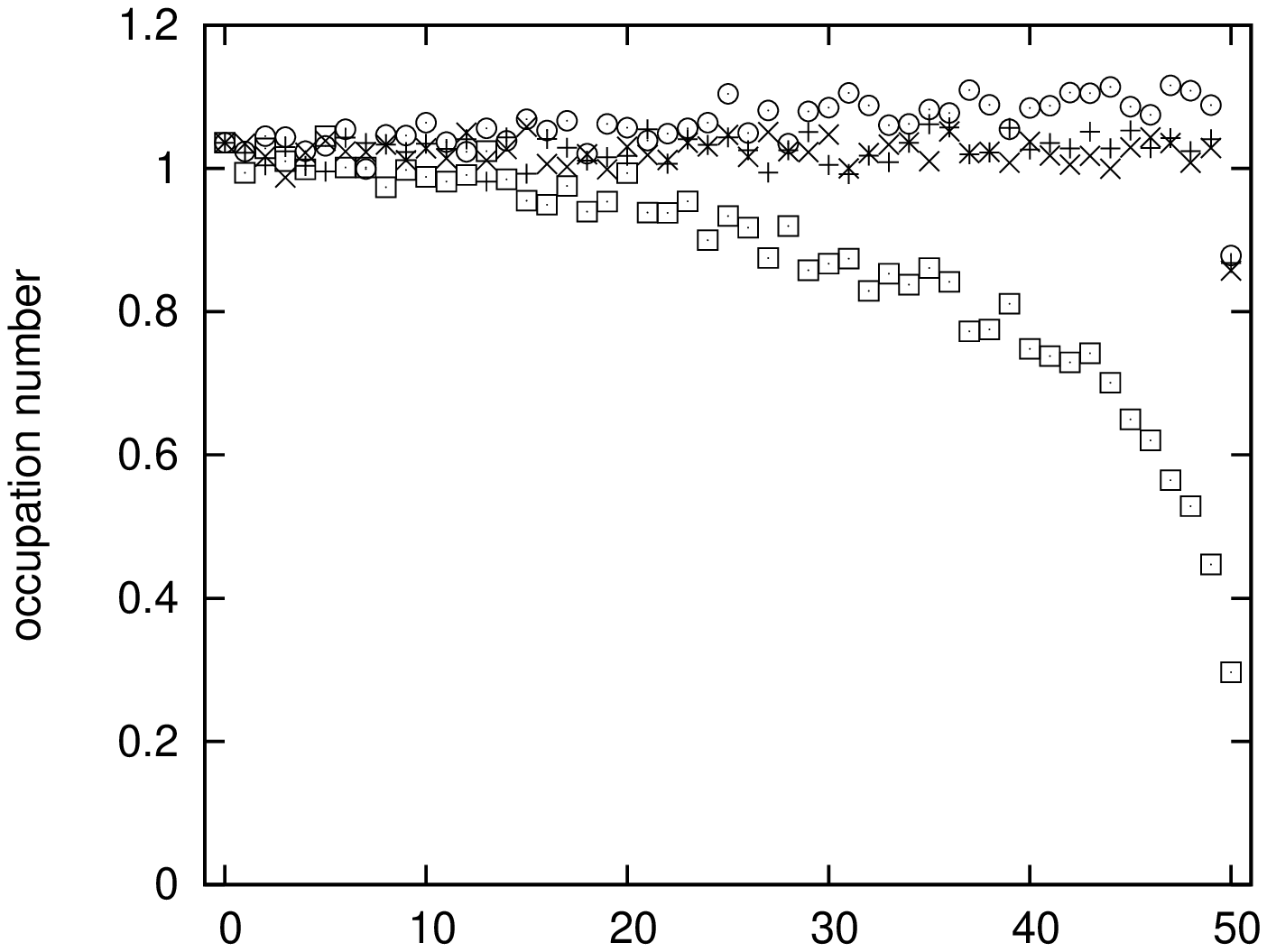}}}
}
\put(90,300)
{
\resizebox{6.5cm}{!}{\rotatebox{-90}{\includegraphics{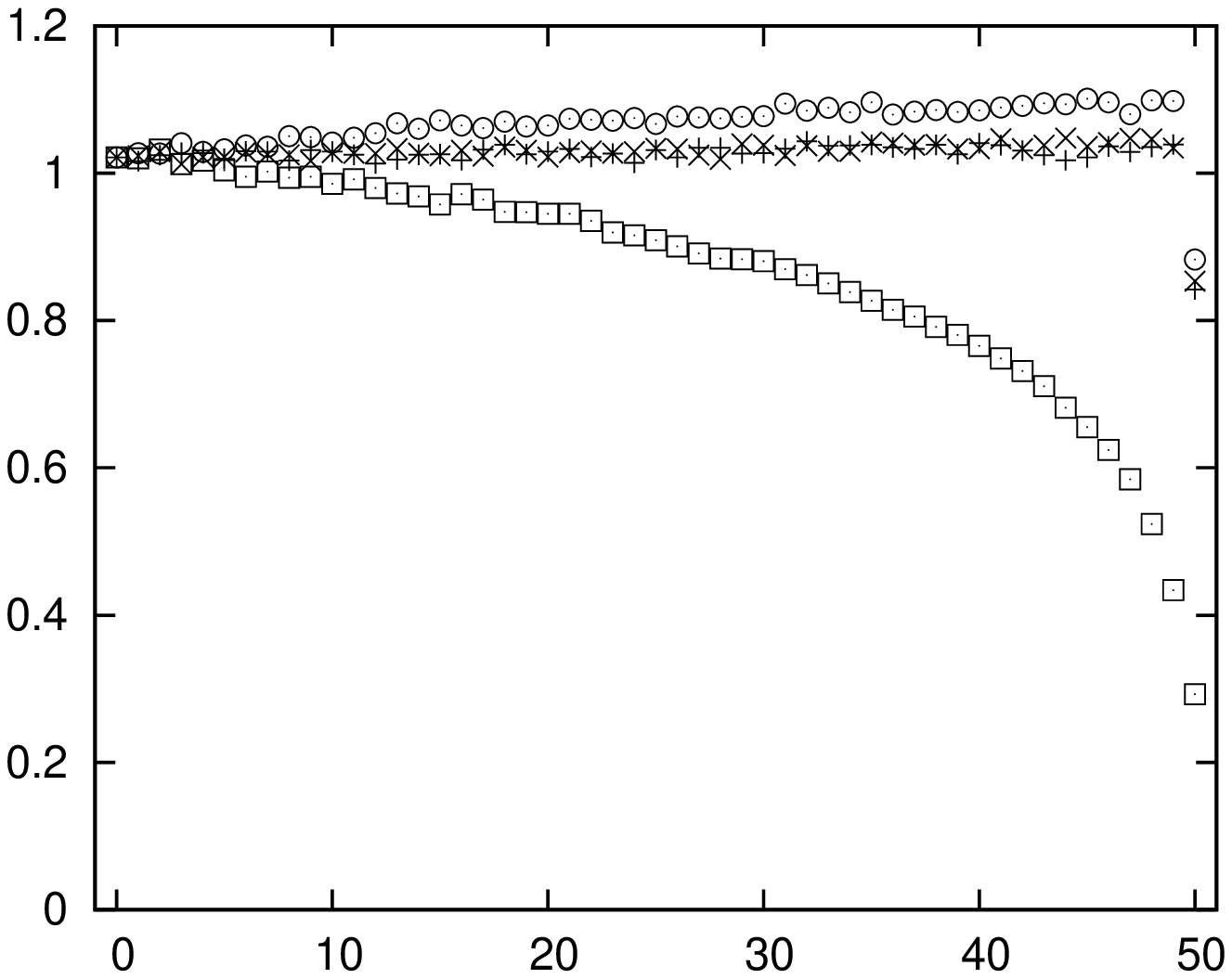}}}
}
\put(260,300)
{
\resizebox{6.5cm}{!}{\rotatebox{-90}{\includegraphics{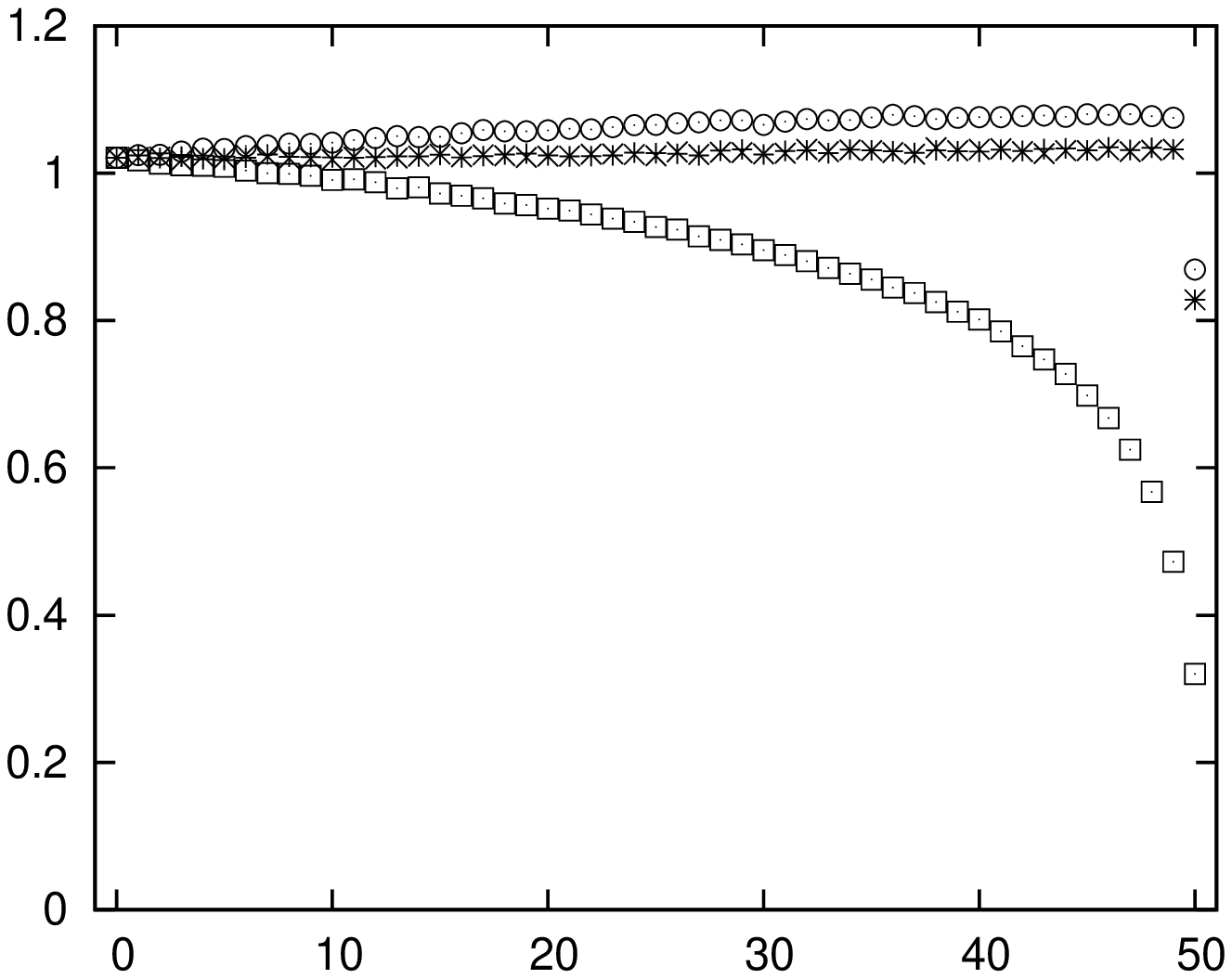}}}
}
\put(-80,440)
{
\resizebox{6.5cm}{!}{\rotatebox{-90}{\includegraphics{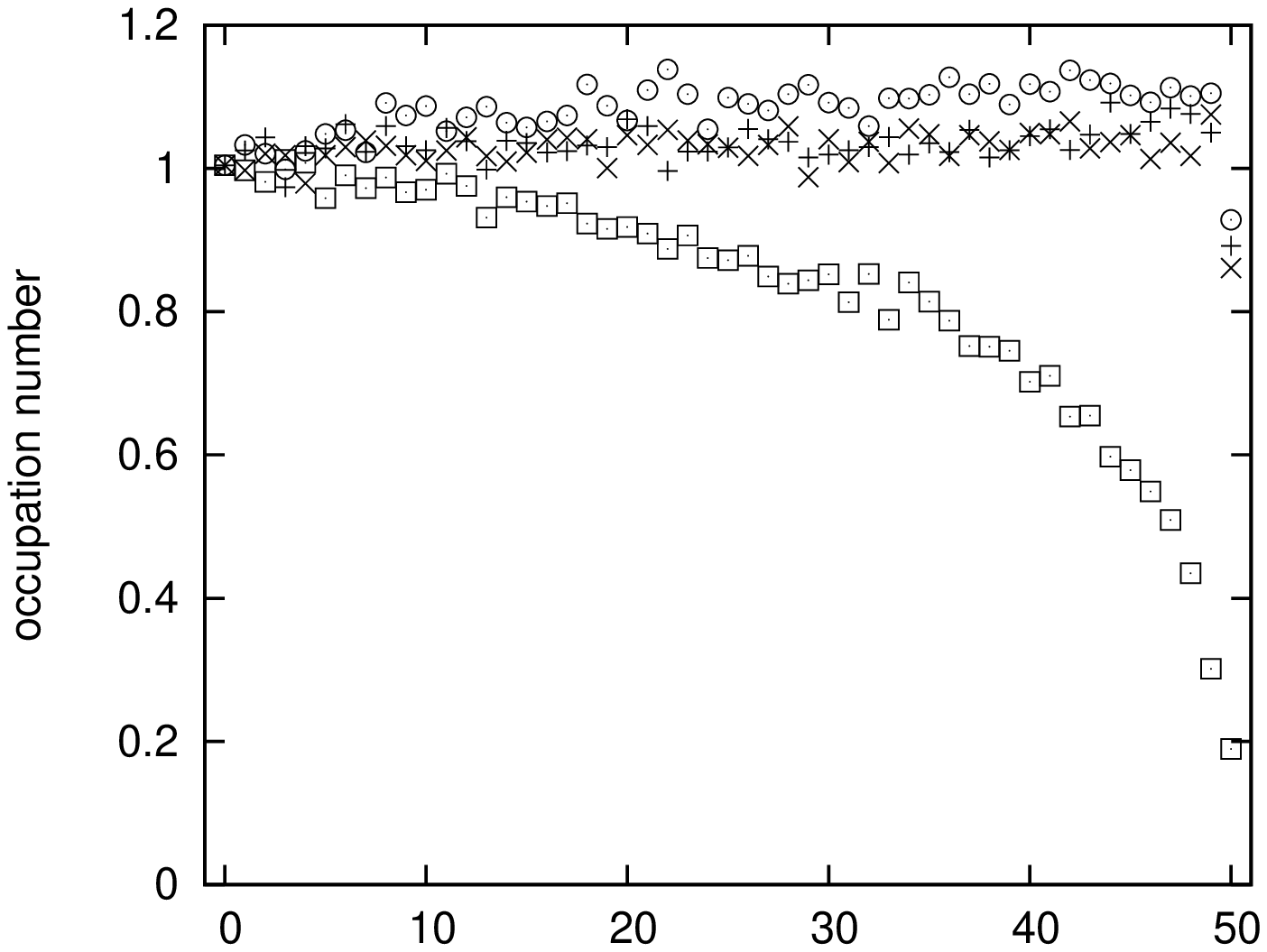}}}
}
\put(90,440)
{
\resizebox{6.5cm}{!}{\rotatebox{-90}{\includegraphics{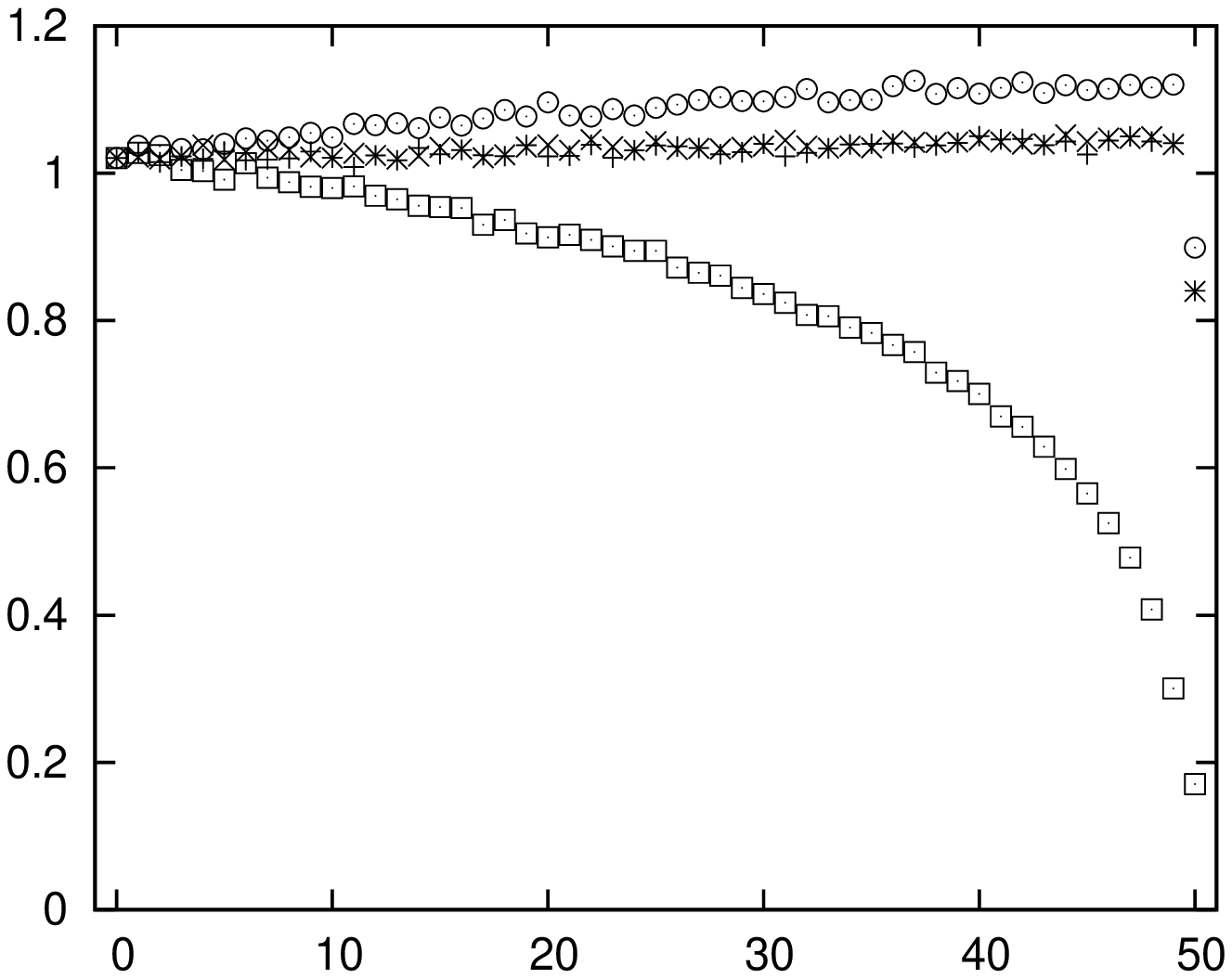}}}
}
\put(260,440)
{
\resizebox{6.5cm}{!}{\rotatebox{-90}{\includegraphics{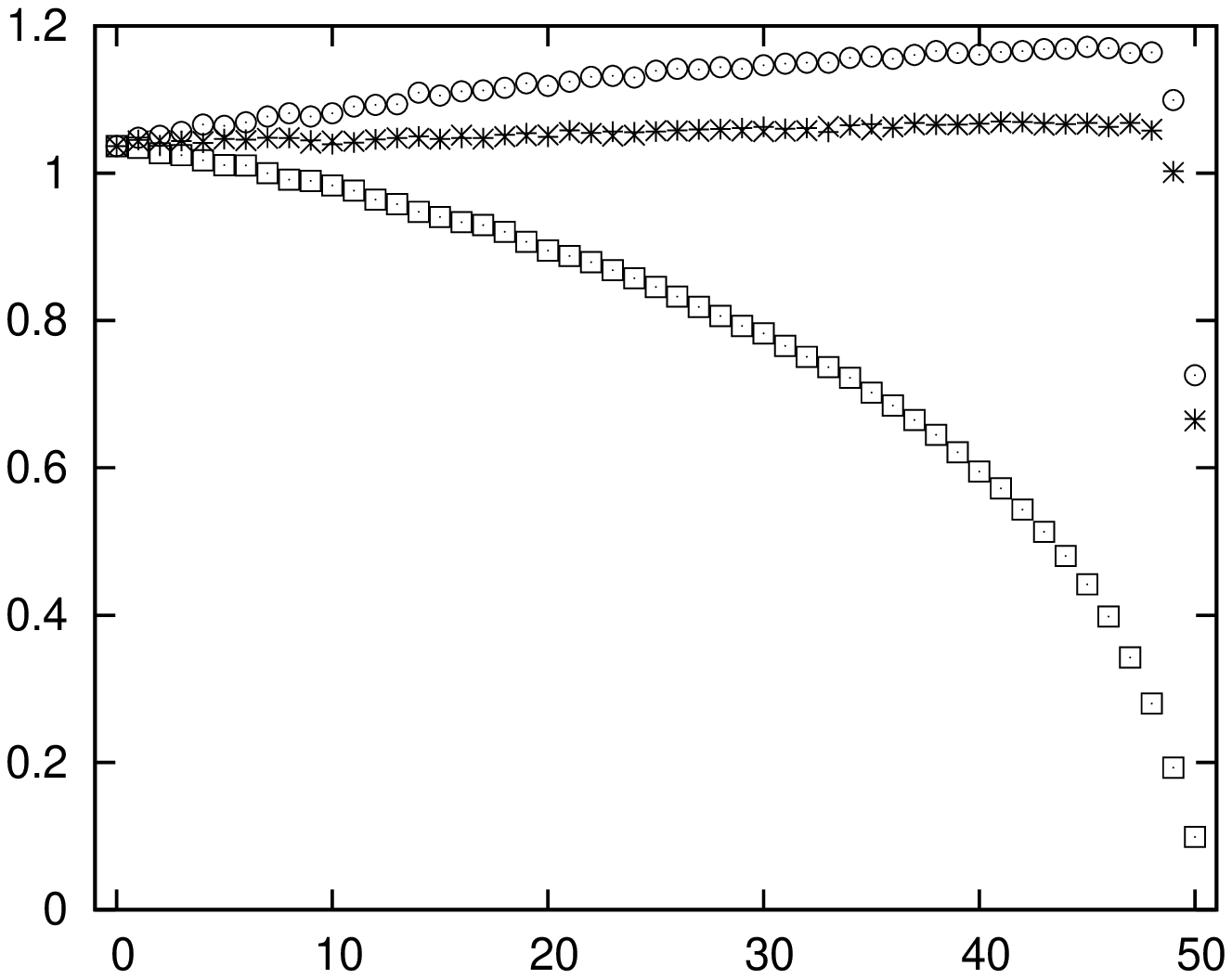}}}
}
\end{picture}
\caption{Stationary occupation number vs. distance from the center.  
The symbols $+$, $\times$, $\circ$, and $\square$
refer respectively to upward, downward, leftward, and rightward 
sites with respect to the center of the lattice. 
Plots refer to the case $N=100,1000,10000$ from the left to the 
right and $T=0,1,5$ from the bottom to the top.}
\label{f:quattro}
\end{center}
\end{figure}

\begin{figure}[htbp]
\begin{center}
\begin{picture}(500,400)
\put(-80,160)
{
\resizebox{6.5cm}{!}{\rotatebox{-90}{\includegraphics{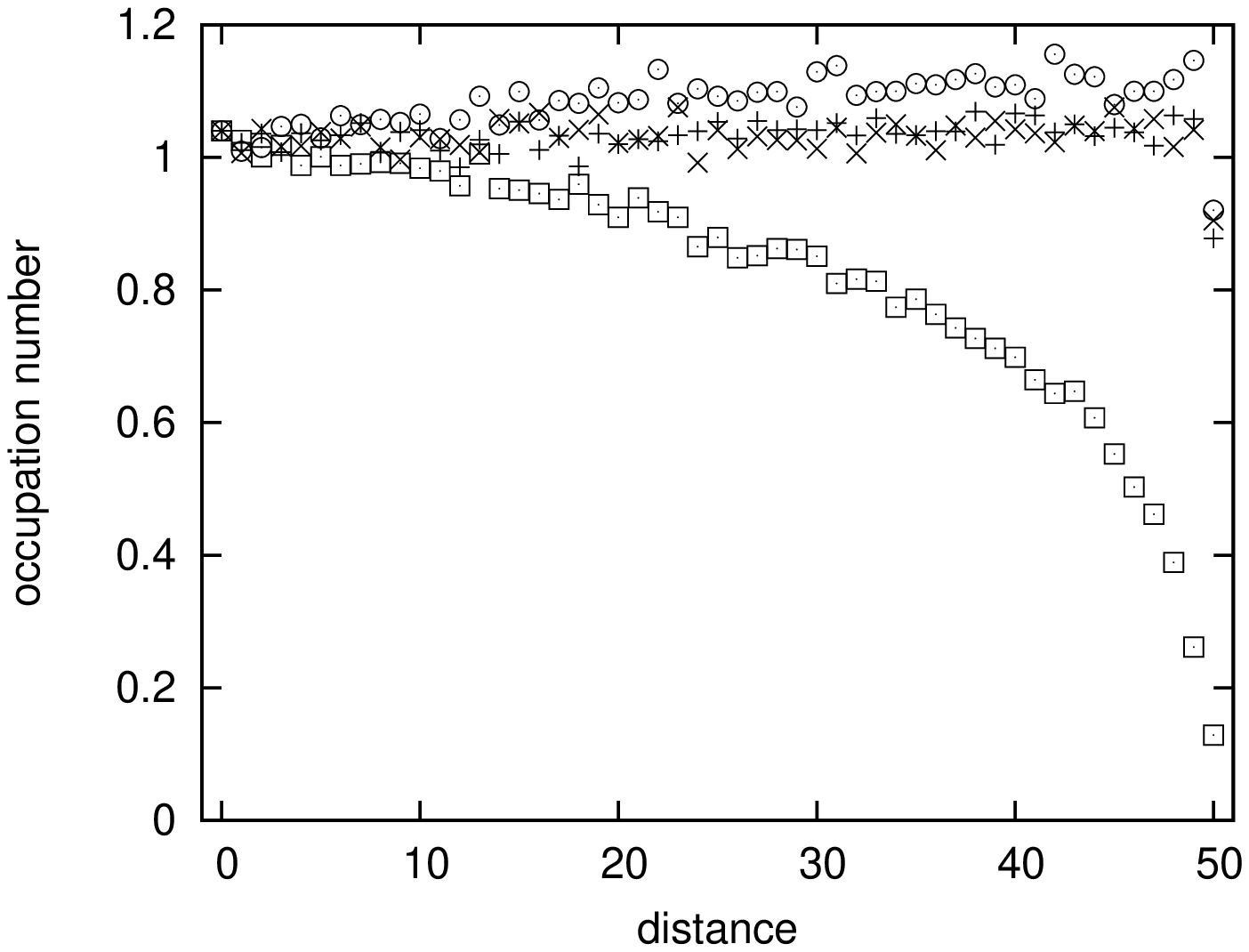}}}
}
\put(90,160)
{
\resizebox{6.5cm}{!}{\rotatebox{-90}{\includegraphics{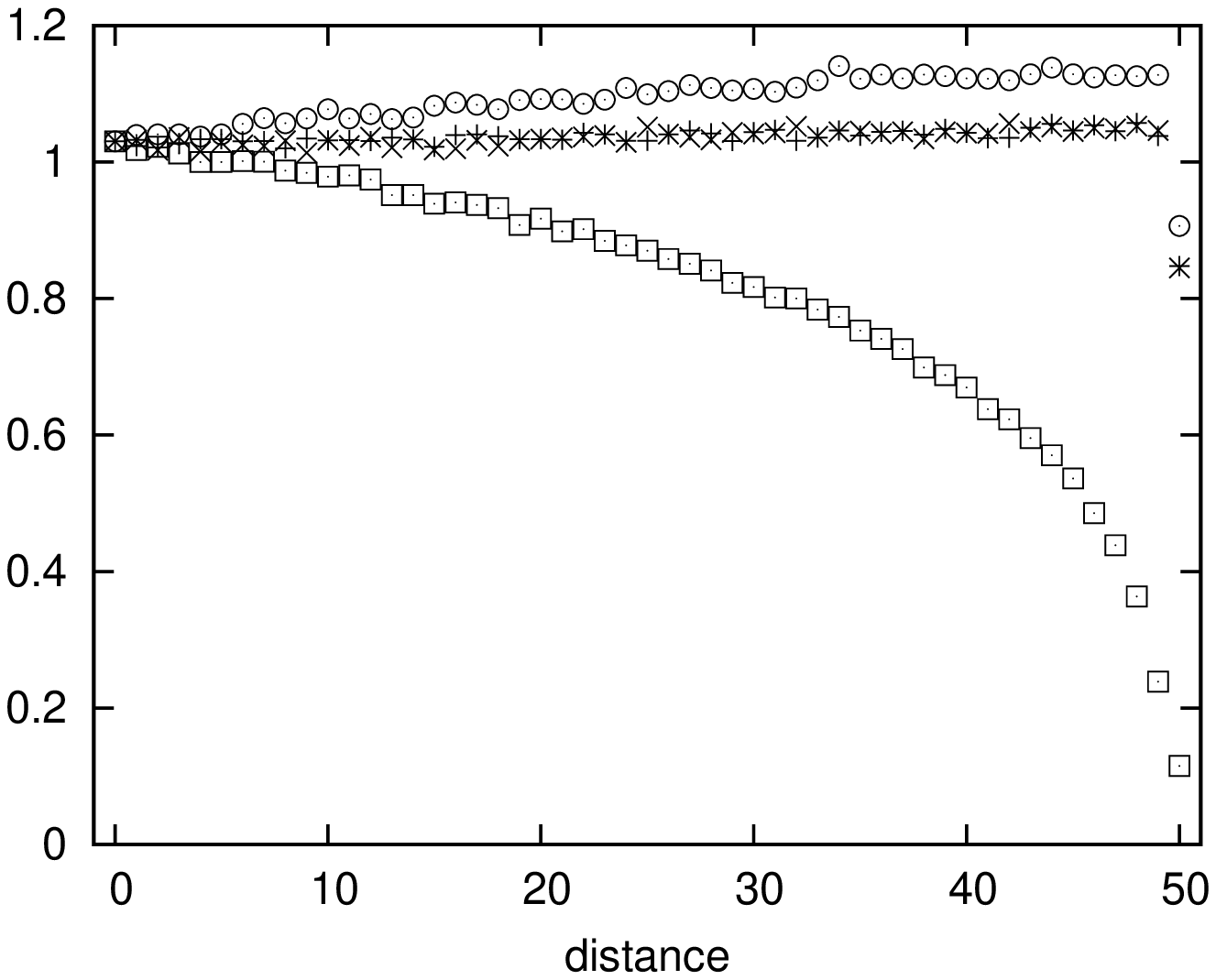}}}
}
\put(260,160)
{
\resizebox{6.5cm}{!}{\rotatebox{-90}{\includegraphics{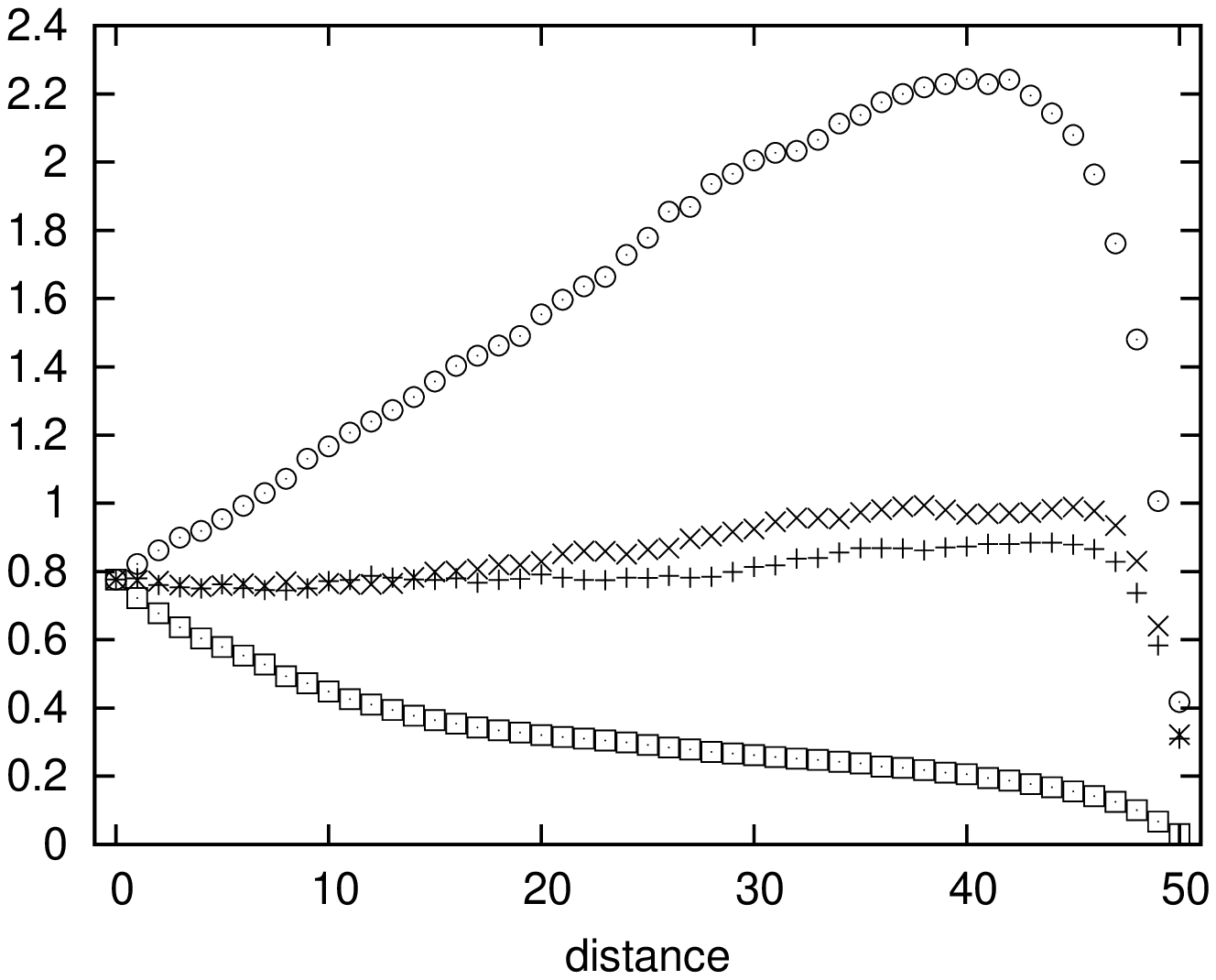}}}
}
\put(-80,300)
{
\resizebox{6.5cm}{!}{\rotatebox{-90}{\includegraphics{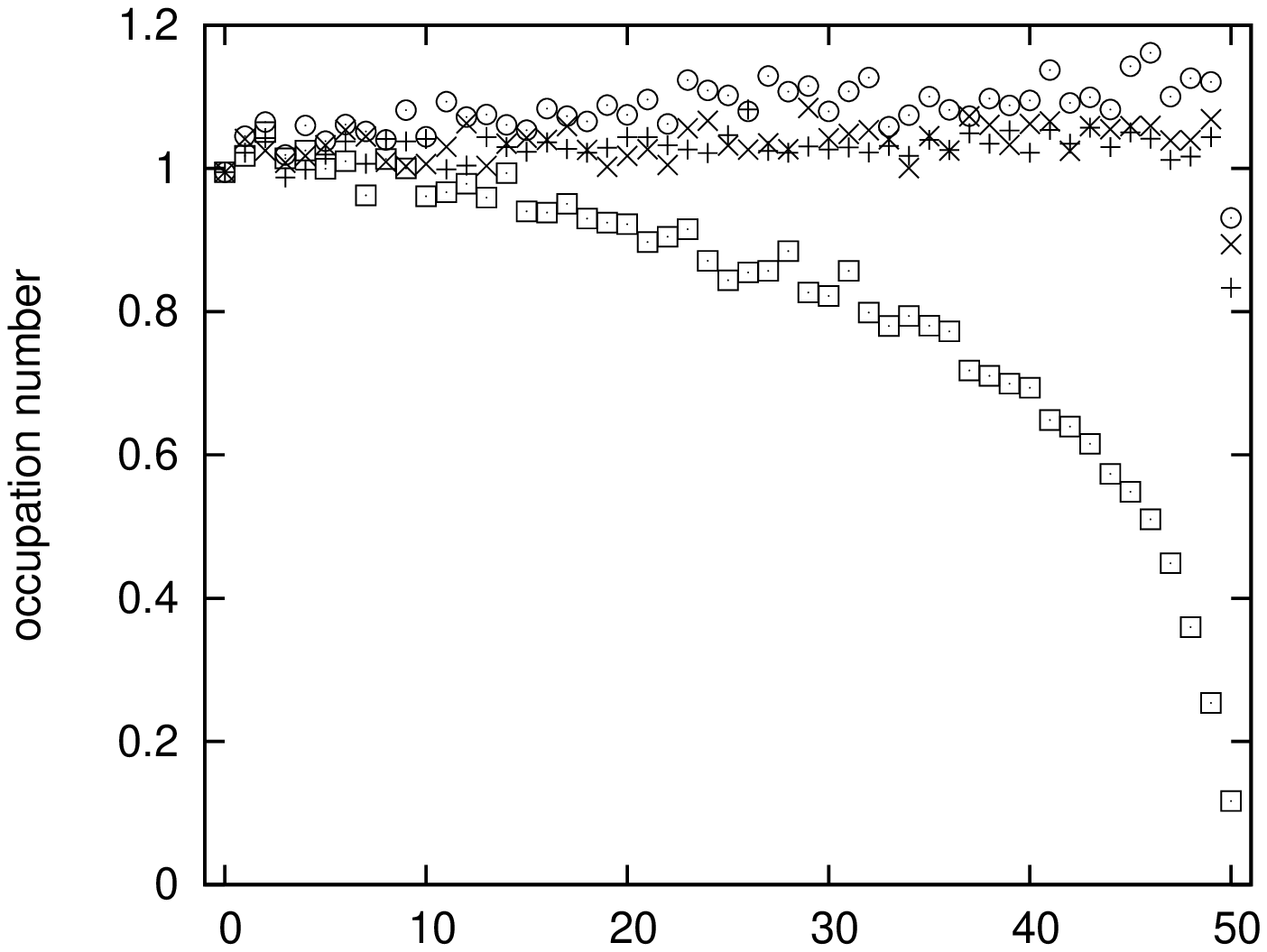}}}
}
\put(90,300)
{
\resizebox{6.5cm}{!}{\rotatebox{-90}{\includegraphics{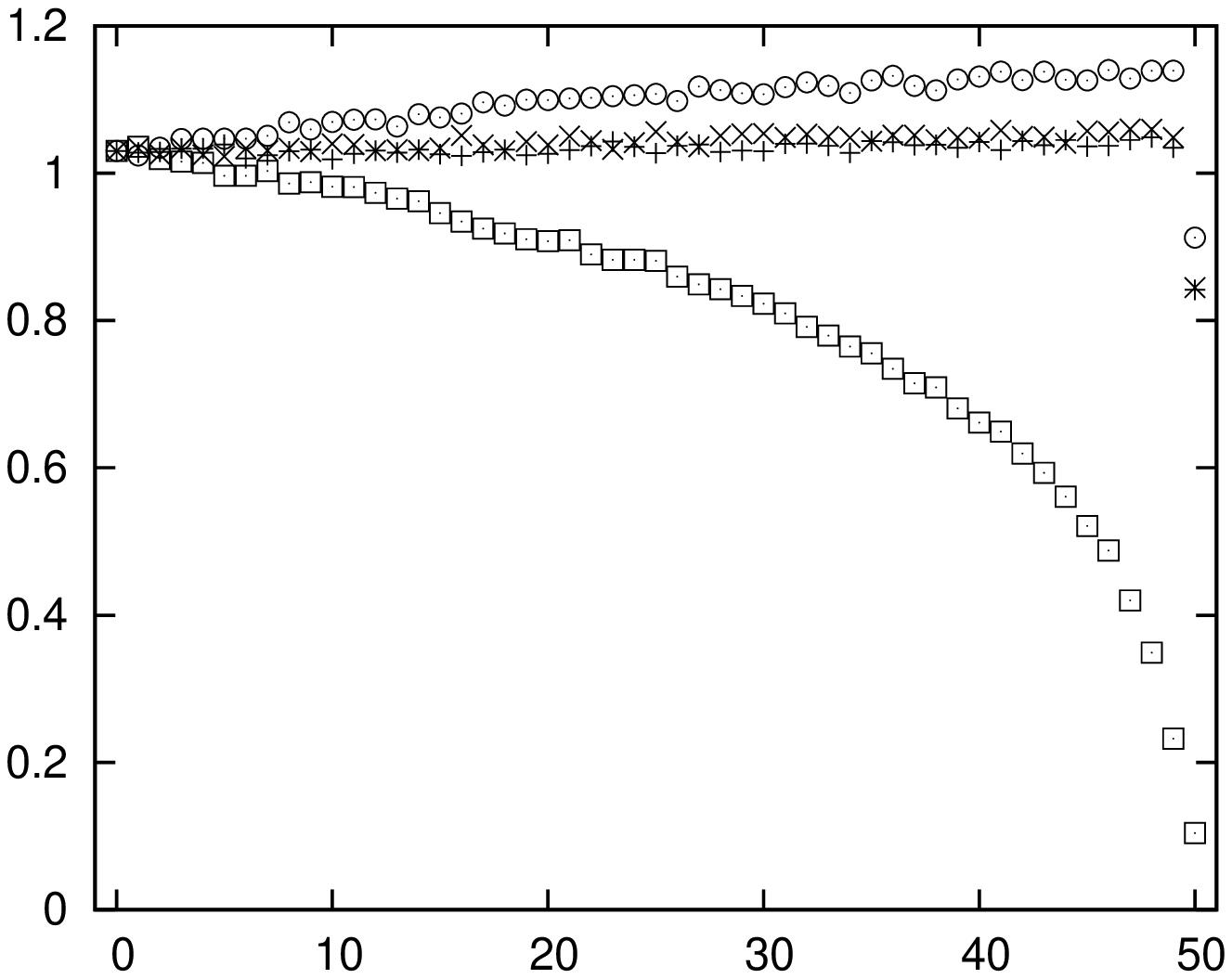}}}
}
\put(260,300)
{
\resizebox{6.5cm}{!}{\rotatebox{-90}{\includegraphics{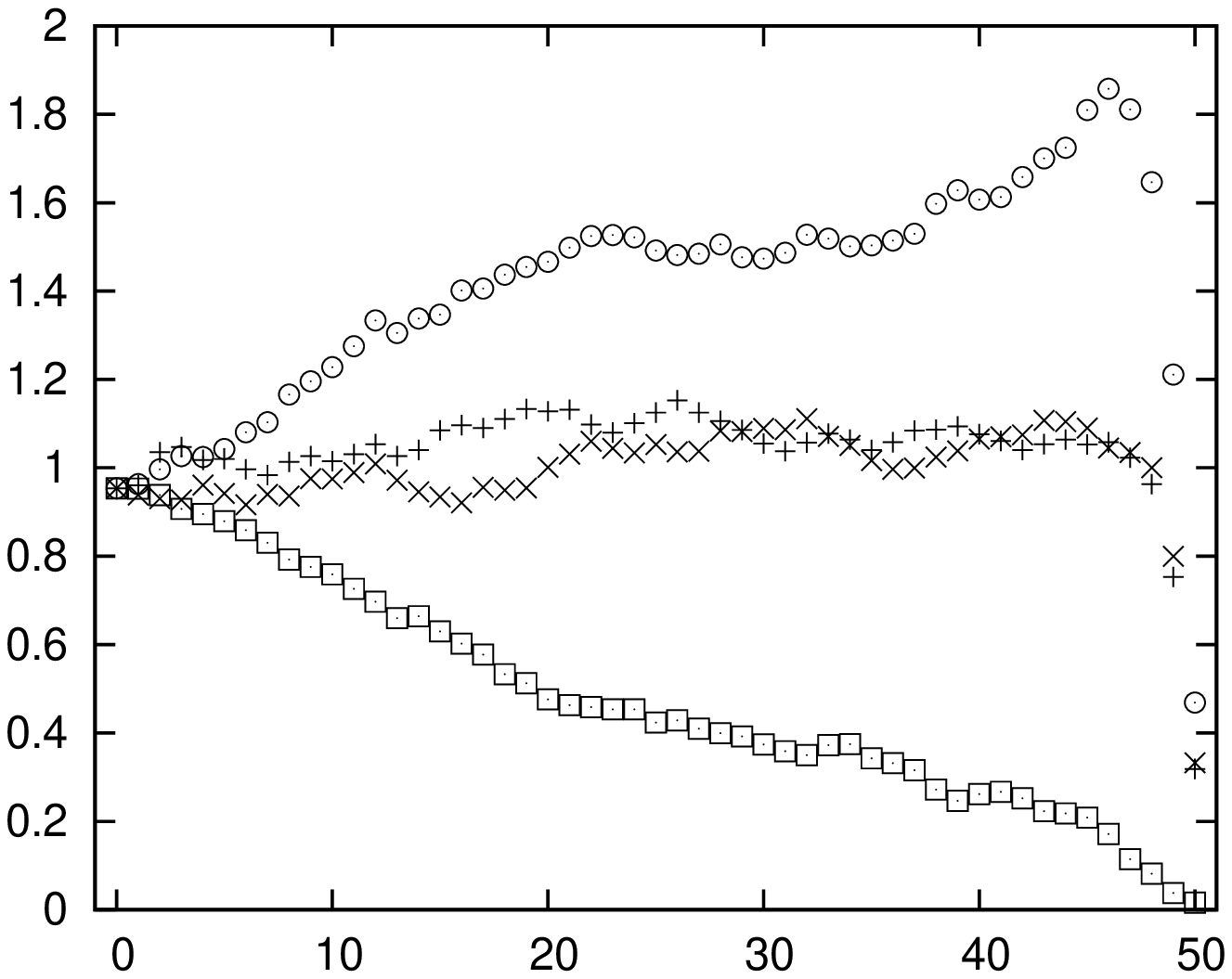}}}
}
\end{picture}
\caption{Stationary occupation number vs. distance from the center.  
The symbols $+$, $\times$, $\circ$, and $\square$
refer respectively to upward, downward, leftward, and rightward 
sites with respect to the center of the lattice. 
Plots refer to the case $N=100,1000,10000$ from the left to the 
right and $T=30,100$ from the bottom to the top.}
\label{f:cinque}
\end{center}
\end{figure}

We have computed the stationary 
occupation number in the center of the 
corridor $\Lambda$, that is at the site whose coordinates 
are both equal to $\lfloor L/2\rfloor+1$, where 
$\lfloor a\rfloor$ denotes the largest integer smaller than 
the real number $a$. Recall that $L$ is an odd number.
Morover we have computed the stationary occupation number 
associated with all the sites of the corridor along the 
two straight lines parallel to the coordinate axes and passing 
through the center of the corridor itself. 

Data  in Figures~\ref{f:quattro} and Figure \ref{f:cinque} 
are plotted as a function of the distance from the center of the corridor. 
We have reported on the horizontal and vertical axes the distance 
of the site from the center of the corridor and the corresponding 
stationary occupation number, respectively. 
The symbols $+$, $\times$, $\circ$, and $\square$
refer respectively to upward, downward, leftward, and rightward 
sites with respect to the center of the lattice. 
The exit of the corridor is to the left of the center of the corridor 
at distance $\lfloor L/2\rfloor+1=51$. 

In all the pictures, we note that the upward and downward behaviors are 
similar; indeed we do not expect any vertical asymmetry. On the other 
and the rightward and leftward behaviors are different with respect 
to each other and, also, with respect to the vertical behavior. 
This is quite intuitive since the exit has been placed on the 
horizontal line passing through the center of the corridor.

In all the pictures it is clearly present a {\em depletion effect} 
in the rightward direction. This was obviously expected. 
This effect does not depend on $N$, while it strongly depends on $T$. 
Particularly,  we note that this gets more and more important with 
increasing $T$. 
It is also worth nothing that the horizontal leftward behavior 
departs from the vertical one at large threshold $T$.

Fixed $T=0,1,5$, the three corresponding graphs at $N=100,1000,10000$ are 
very similar among each others, see figure~\ref{f:quattro}.
On the other hand at $T=30,100$ the behavior at $N=10000$ is somehow 
singular, see figure~\ref{f:cinque}.
This suggests that for large threshold $T$ and number of individuals 
$N$ the system reaches a stationary state different from the one characterizing
the small threshold and/or small number of pedestrian regimes. 
This remark sounds reasonable also in view of the 
outgoing flux behaviors recorded in figure~\ref{f:due}.

A possible explanation for this regime is that
pile of individuals are formed and the outgoing flux is controlled 
by the pile dynamics, which is typically much slower than the single
partcle one.

\begin{figure}[htbp]
\begin{center}
\begin{picture}(500,400)
\put(-80,160)
{
\resizebox{6.5cm}{!}{\rotatebox{-90}{\includegraphics{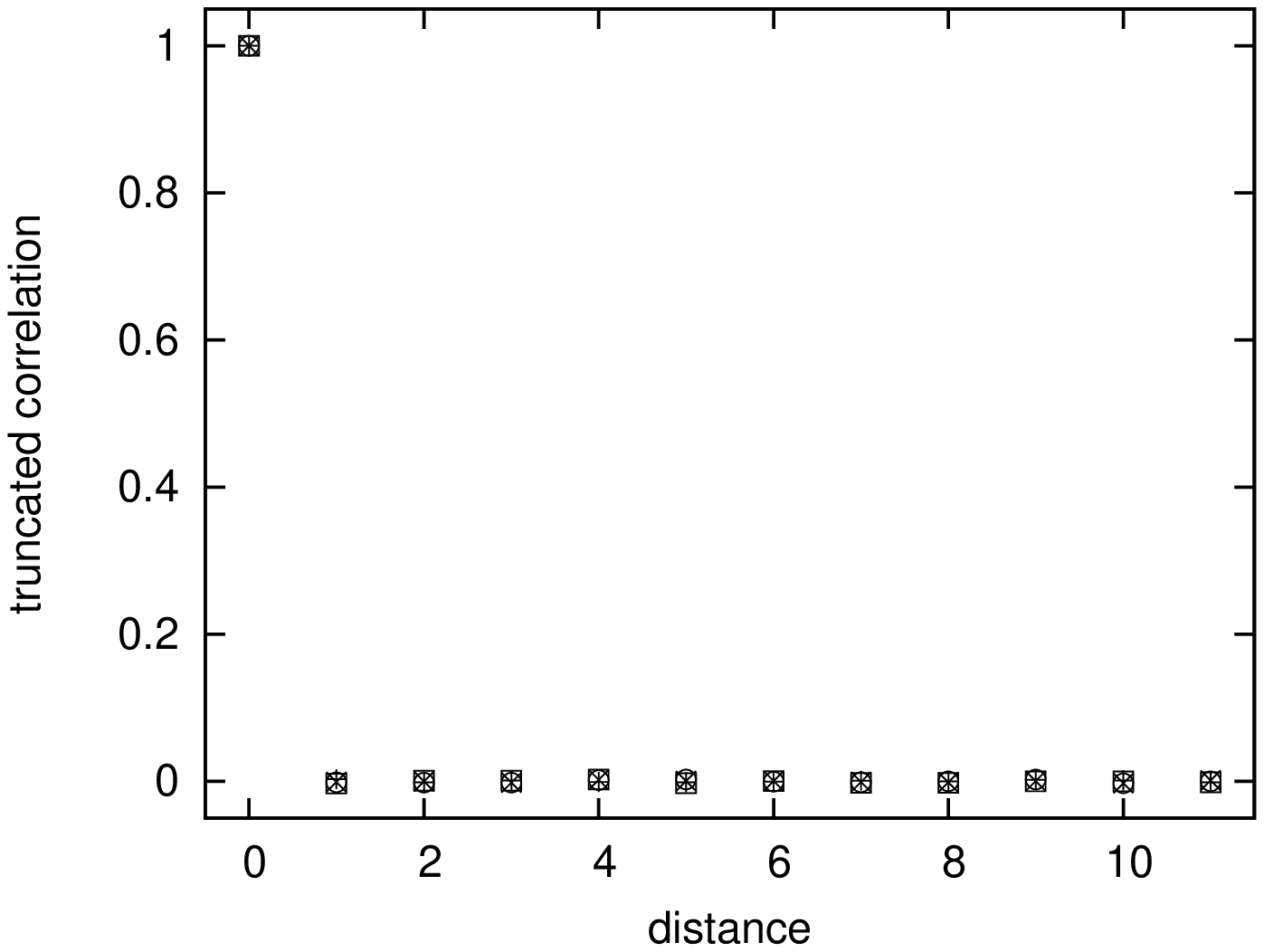}}}
}
\put(90,160)
{
\resizebox{6.5cm}{!}{\rotatebox{-90}{\includegraphics{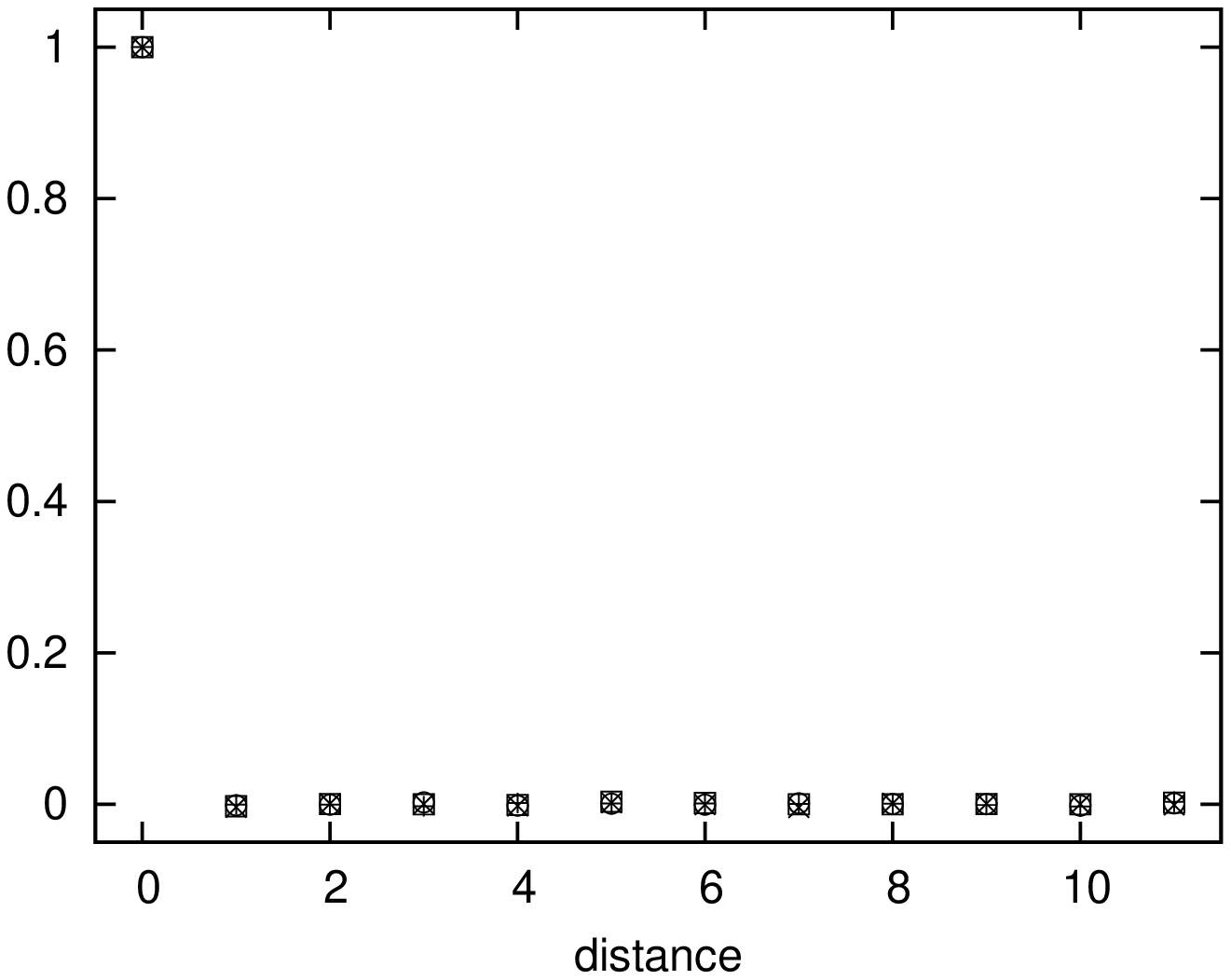}}}
}
\put(260,160)
{
\resizebox{6.5cm}{!}{\rotatebox{-90}{\includegraphics{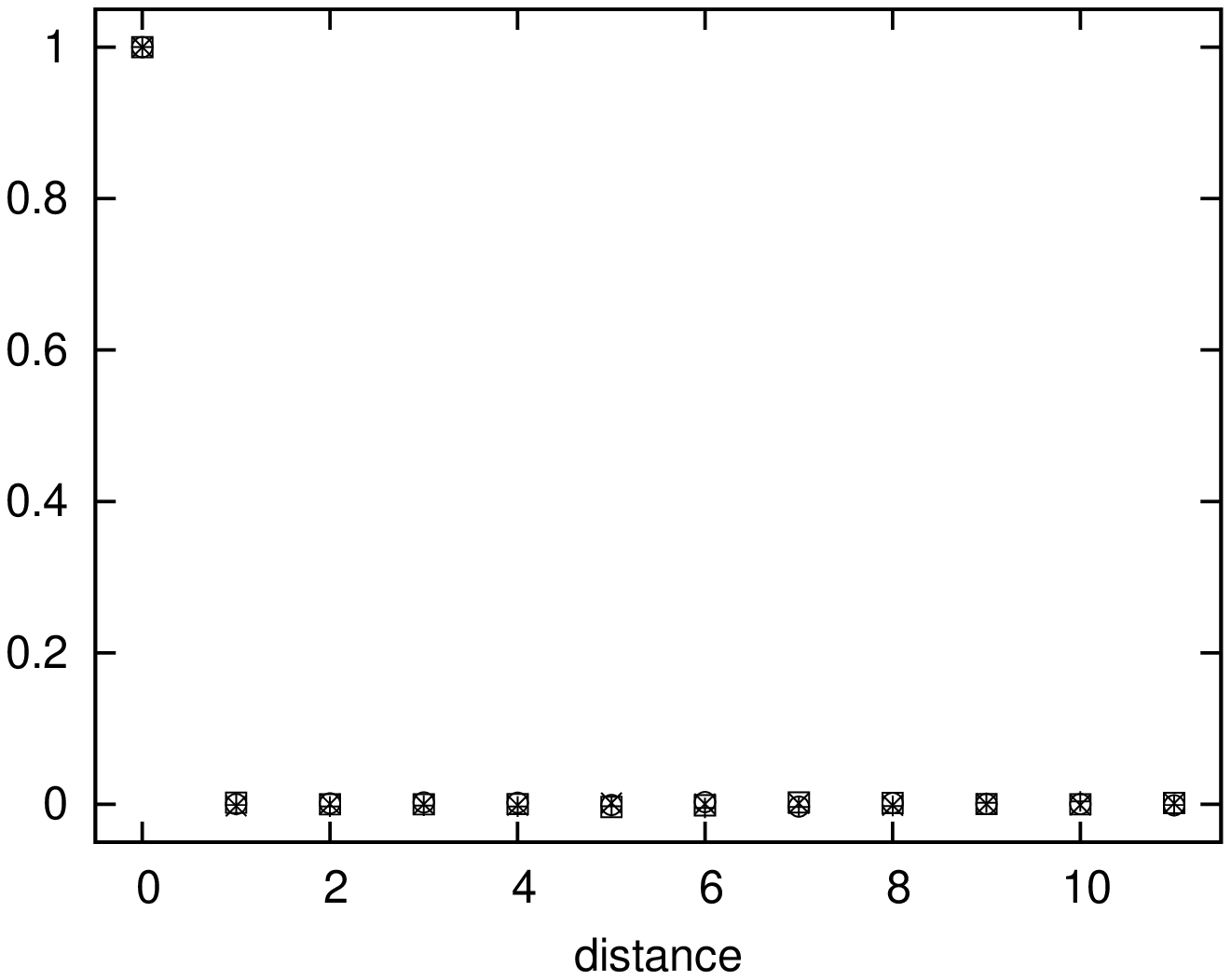}}}
}
\put(-80,300)
{
\resizebox{6.5cm}{!}{\rotatebox{-90}{\includegraphics{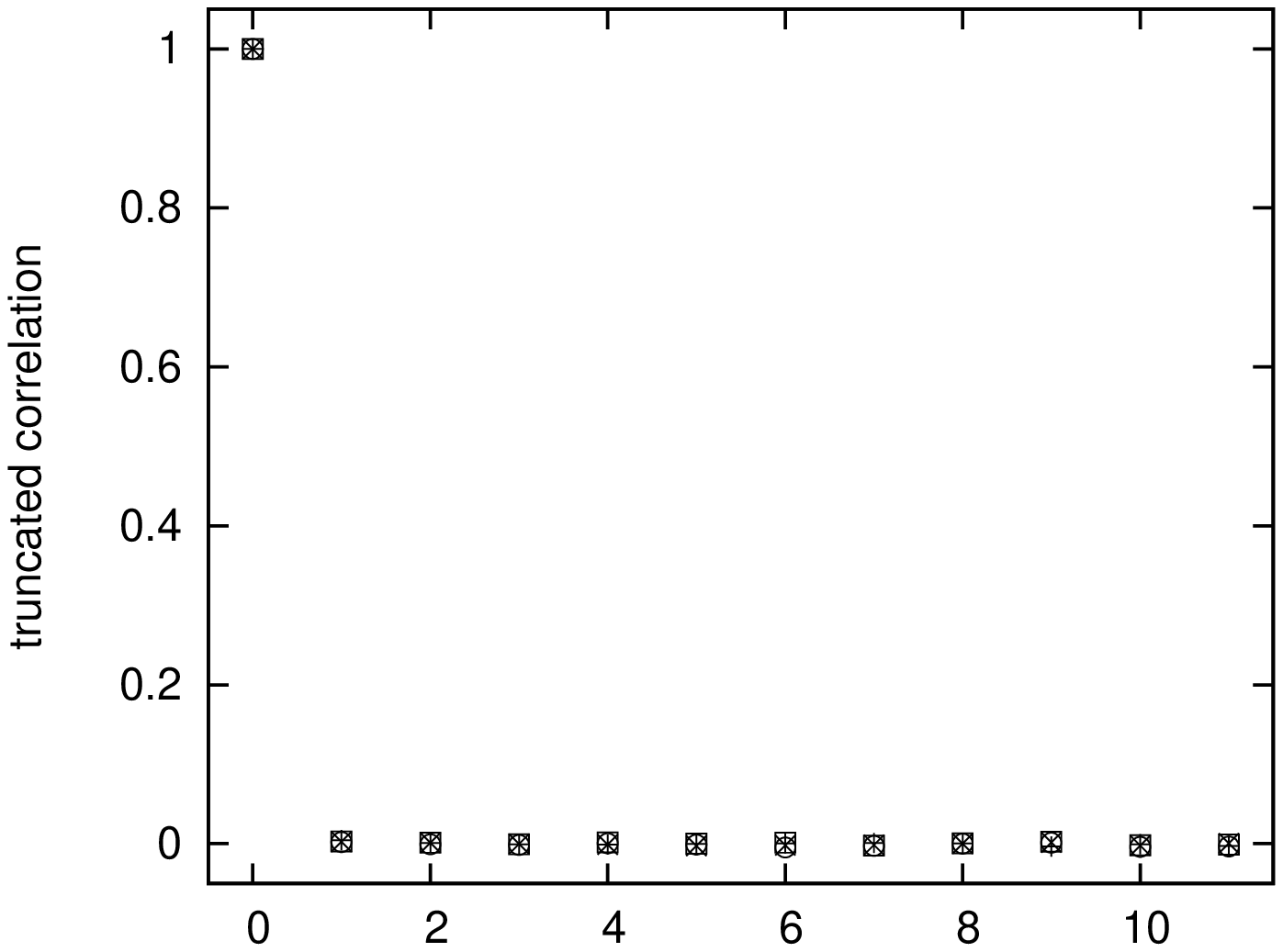}}}
}
\put(90,300)
{
\resizebox{6.5cm}{!}{\rotatebox{-90}{\includegraphics{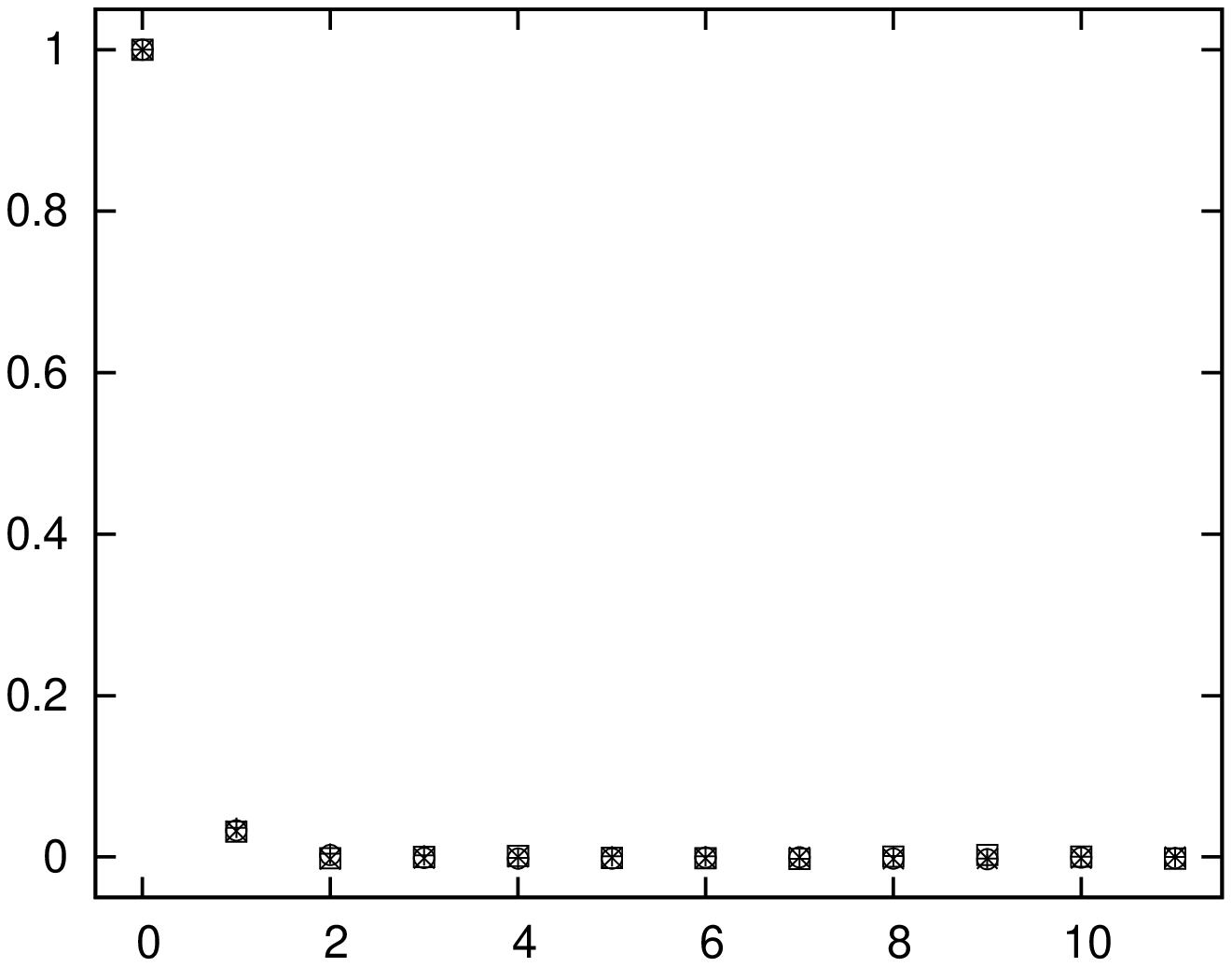}}}
}
\put(260,300)
{
\resizebox{6.5cm}{!}{\rotatebox{-90}{\includegraphics{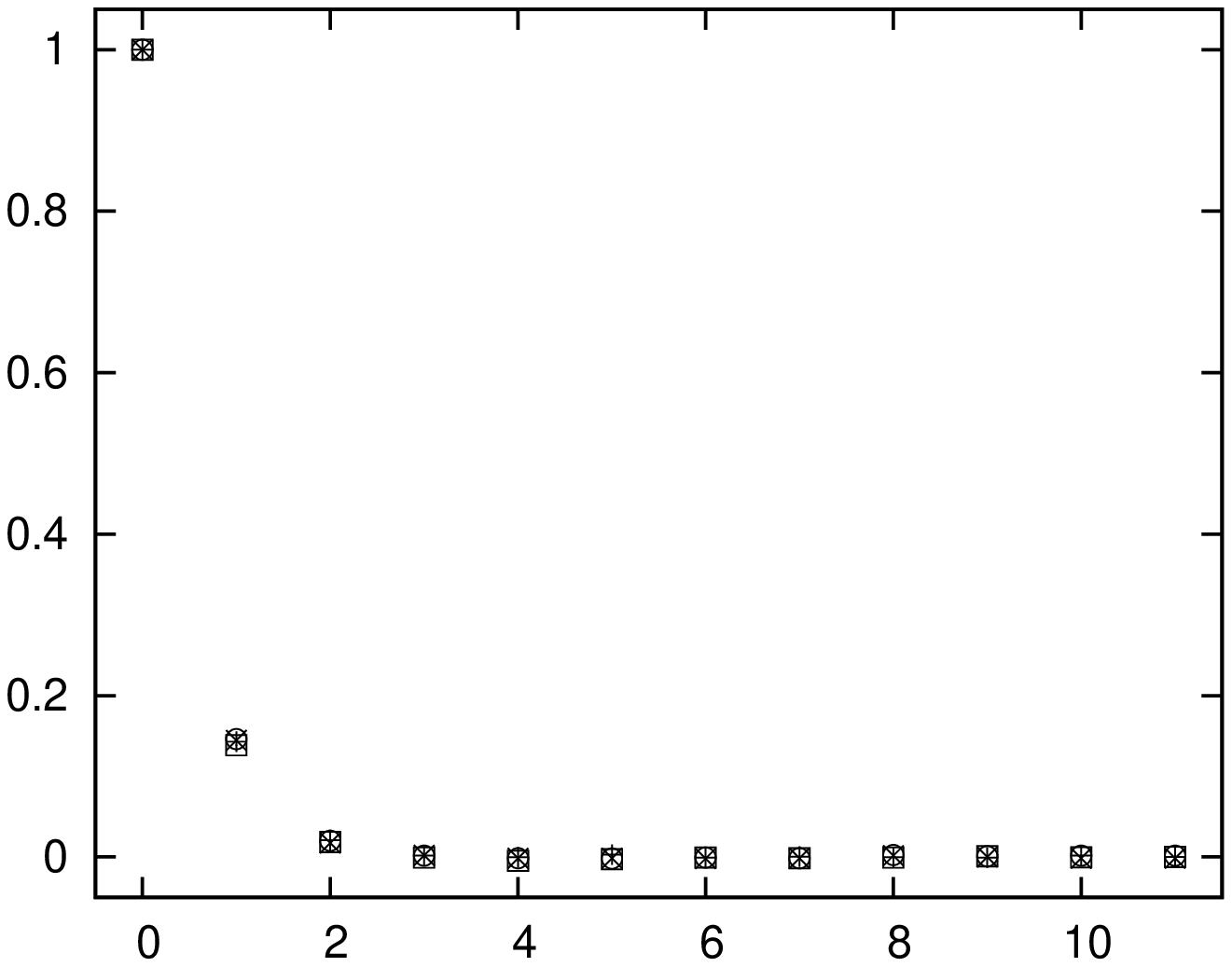}}}
}
\put(-80,440)
{
\resizebox{6.5cm}{!}{\rotatebox{-90}{\includegraphics{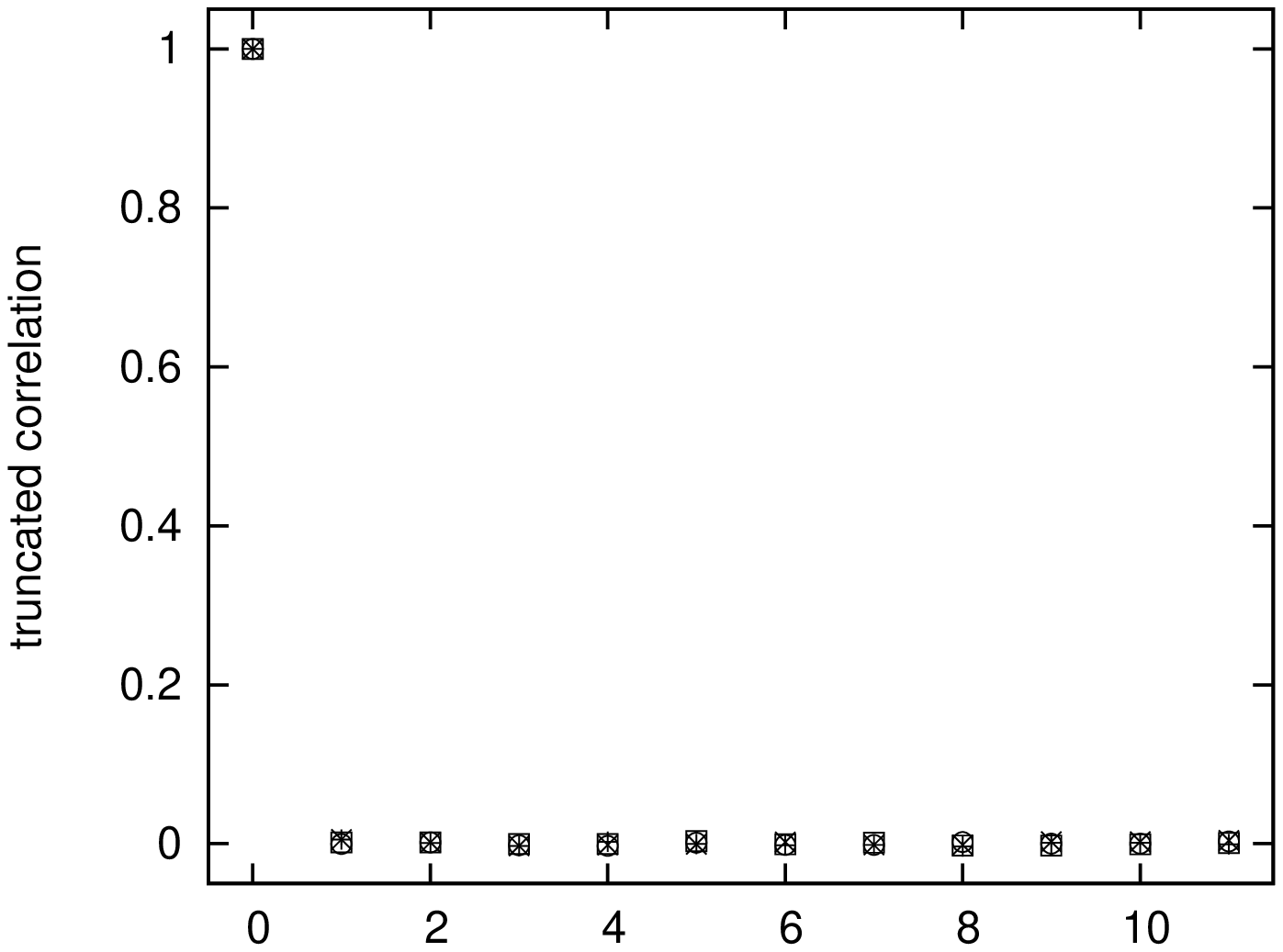}}}
}
\put(90,440)
{
\resizebox{6.5cm}{!}{\rotatebox{-90}{\includegraphics{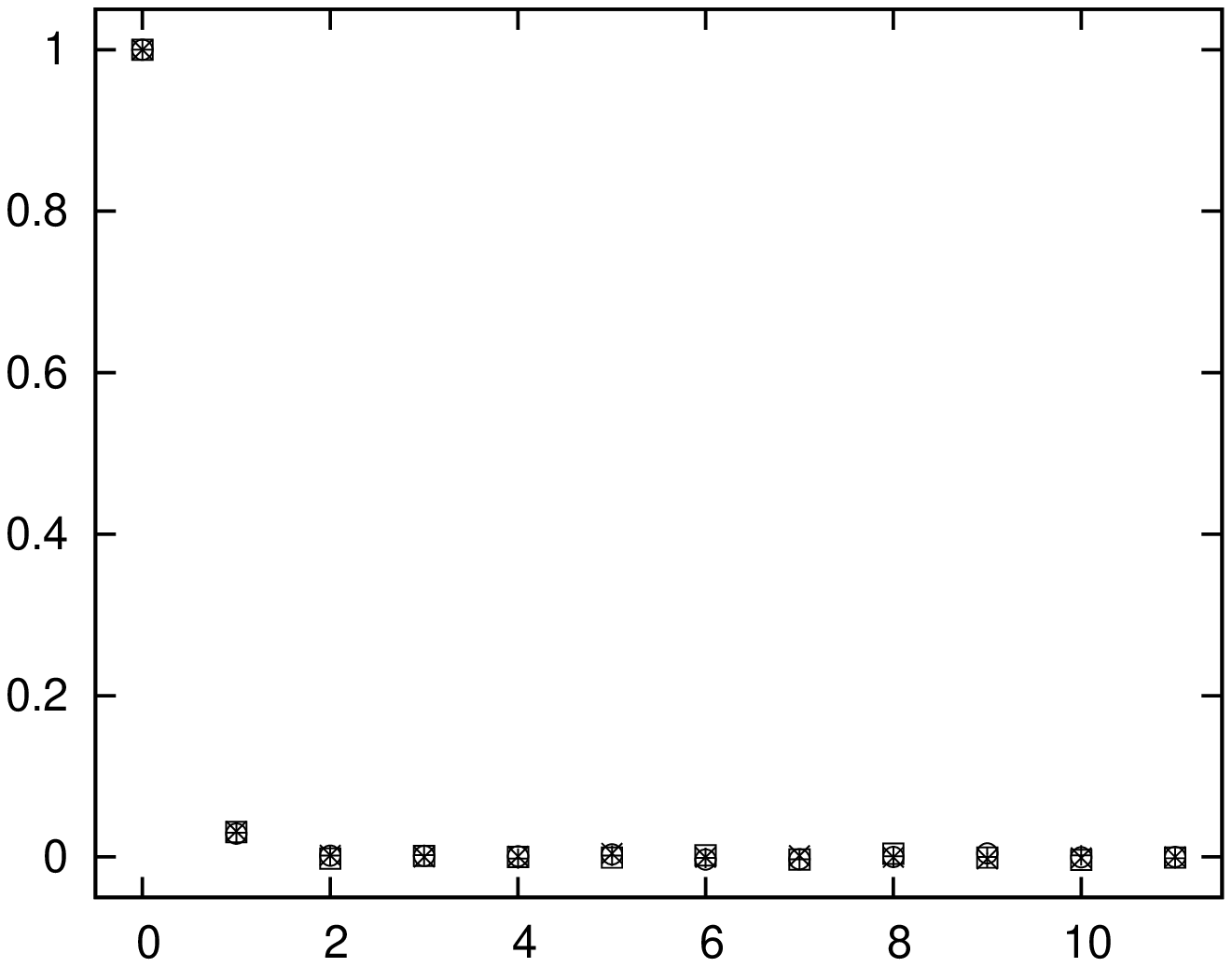}}}
}
\put(260,440)
{
\resizebox{6.5cm}{!}{\rotatebox{-90}{\includegraphics{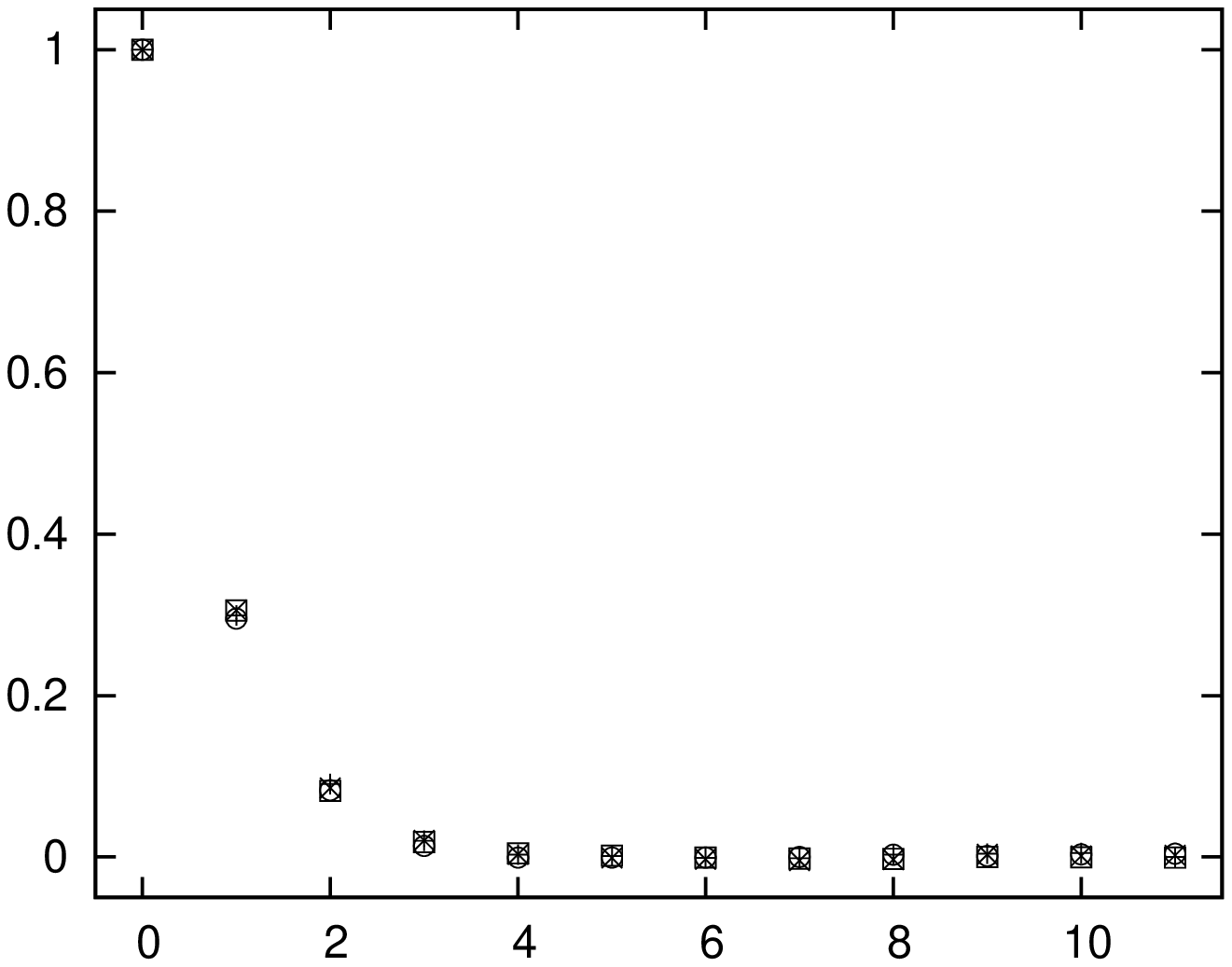}}}
}
\end{picture}
\caption{Stationary truncated correlations vs. distance from the center.  
The symbols $+$, $\times$, $\circ$, and $\square$
refer respectively to upward, downward, leftward, and rightward 
sites with respect to the center of the lattice. 
Plots refer to the case $N=100,1000,10000$ from the left to the 
right and $T=0,5,100$ from the bottom to the top.}
\label{f:sei}
\end{center}
\end{figure}

\par\noindent
\subsection{Stationary correlations}
\label{s:correlazioni}
\par\noindent
Another relevant quantity we have studied at equilibrium 
is the correlation between the occupation 
number random variables.
Recall that
$\langle\cdot\rangle$ denotes the 
stationary average of a random variable associated with the system.
For any $x,y\in\Lambda$ 
we call {\em truncated correlation} of the occupation numbers at sites 
$x$ and $y$ the quantity
\begin{displaymath}
\langle 
[u(x)-\langle u(x)\rangle]
[u(y)-\langle u(y)\rangle]
\rangle
=
\langle u(x)u(y) \rangle
-
\langle u(x)\rangle
\langle u(y)\rangle
\end{displaymath}
that is to say 
the covariance between those random variables. Note that for $x=y$ 
we simply get the variance of the random variable $u(x)$.

In Figure~\ref{f:sei}, we have plotted the normalized truncated correlations 
of the occupation numbers of the site in the center of the corridor 
and a site on the vertical and on the horizontal straight line 
parallel to the coordinate axes passing through the center of the 
corridor itself. 
In all the figures, the truncated correlations have been normalized 
times the covariance of the occupation number at the center 
of the corridor. 

The information in Figure~\ref{f:sei} is represented 
as a function of the distance from the center of the corridor. 
We show there on the horizontal and vertical axes the distance 
of the site from the center of the corridor and the corresponding 
normalized truncated correlation, respectively. 
The symbols $+$, $\times$, $\circ$, and $\square$
refer respectively to upward, downward, leftward, and rightward 
sites with respect to the center of the lattice. 

The relevant remark at this point is that 
at $T=0$ no correlation is detected, while at $T>0$ 
nonzero correlations start to occur at  sufficiently large $N$. 

\section{Discussions}\label{s:discussion}

In this section, we only wish to emphasize a few basic aspects concerning the
dynamics of pedestrians within regions with no visibility that we discover with
our modeling of buddying:
\begin{itemize}
\item[--]
In \cite{Simo}, the authors present an experiment whose purpose was to
study evacuees¿ exit selection under different behavioral objectives. The
experiment was conducted in a corridor with two exits located asymmetrically.
This geometry was used to make most participants face a nontrivial decision on
which exit to use. 
The statistical approach used in \cite{Simo}  seems to indicate that the
individuals of an evacuating crowd may not be able to make optimal decisions
when assessing the fastest exit to evacuate. In their case, the evacuation
(egress) time of the whole crowd turned out to be shorter when the evacuees
behave egoistically instead of behaving cooperatively. This is rather
intriguing and counter intuitive fact, and it is very much in the spirit of the
effect of the threshold $T$ we observed in section \ref{s:risultati}.
\item[--]
Note that for low densities the buddying mechanism increase
the outgoing flux: compare points and straight line
in the pictures (straight line is essentially the not--buddying case),
while at large densities the scenario is dramatic: 
isolated individuals may turn
to have a bigger escape chance than a large group around a leader [behavior
recommended by standard manuals on evacuation strategies, see e.g.
\cite{NIBHV}, p. 122.]. This suggests that evacuation strategies should not
rely too much on the presence of a leader.
\item[--]
Quite probably, in order to decide what is the best strategy for a given
scenario {\em cooperation} (grouping, buddying, etc.) or {\em selfishness}
(walking away from groups)  one would need to explore also basic aspects
connected to  the prisoner's dilemma, or to a broader context  like stochastic
game theory \cite{Szilagyi}.
\end{itemize}

\section*{Acknowledgments}
AM thanks Michael B\"ohm (Bremen) and Martin Klein (DGMR, Arnhem) for fruitful discussions on the motion of people inside smoky regions. 
ENMC partially did this work at Eurandom during the
Stochastic Activity Month, February 2012.
ENMC wants to espress his thanks to the organizers, R.\ Fernandez,
R.\ van der Hofstad, and M.\ Heydenreich, for the invitation and to
Eurandom for the kind hospitality. 
ENMC also thanks A.\ Asselah for a useful discussion.

\bibliographystyle{elsarticle-num}
\bibliography{smoke}
\end{document}